\documentclass[prb,aps,amsmath,amssymb,twocolumn,groupedaddress,floats,showpacs,final]{revtex4-1}
\usepackage{graphicx}
\usepackage{dcolumn}
\usepackage{bm}
\usepackage{color}

\newcommand{\MC}{\text MC}

\usepackage{hyperref}
\hypersetup{
  pdftitle={Bold Diagrammatic Monte Carlo technique for frustrated
    spin systems},
  pdfauthor={S. Kulagin, N. Prokof'ev, O. A. Starykh, B. Svistunov,
    and C. N. Varney},
  colorlinks=true,
  linkcolor=blue,
  citecolor=blue,
  pdfpagemode=UseNone
}

\begin{document}

\title{Bold Diagrammatic Monte Carlo technique for frustrated spin systems}

\author{S.~A. Kulagin$^{1,2}$, N. Prokof'ev$^{1,3}$, O.~A. Starykh$^{4}$, B. Svistunov$^{1,3}$, and C.~N. Varney$^{1}$}

\affiliation{$^1$Department of Physics, University of Massachusetts,
  Amherst, Massachusetts 01003, USA\\
$^2$Institute for Nuclear Research of Russian Academy of Sciences,
  117312 Moscow, Russia\\
$^3$Russian Research Center ``Kurchatov Institute",
  123182 Moscow, Russia\\
$^4$Department of Physics and Astronomy, University of
  Utah, Salt Lake City, UT 84112, USA}

\date{\today}
\begin{abstract}
  Using fermionic representation of spin degrees of freedom within the
  Popov-Fedotov approach we develop an algorithm for Monte Carlo
  sampling of skeleton Feynman diagrams for Heisenberg type
  models. Our scheme works without modifications for any dimension of
  space, lattice geometry, and interaction range, i.e. it is suitable
  for dealing with frustrated magnetic systems at finite
  temperature. As a practical application we compute uniform magnetic
  susceptibility of the antiferromagnetic Heisenberg model on the
  triangular lattice and compare our results with the best available
  high-temperature expansions. We also report
  results for the momentum-dependence of the static magnetic susceptibility
  throughout the Brillouin zone.
  \end{abstract}


\maketitle

\section{Introduction}
\label{sec:1}

Properties of geometrically frustrated spin systems in various
dimensions, geometries, and temperature regimes are at the heart of
modern condensed matter physics. Here, {\em frustration} is a technical term
which refers to the presence of competing forces that cannot be
simultaneously satisfied. In numerous quantum antiferromagnets
frustration often has a simple geometric origin. Localized spins on
two- and three-dimensional lattices with triangular motifs, such as
planar triangular antiferromagnet and three-dimensional pyrochlore
antiferromagnets, cannot assume energetically favorable antiparallel
orientation. Of three spins forming a minimal triangle, and
interacting via simple antiferromagnetic pair-wise exchange
interaction, only two can be made antiparallel, leaving the third one
frustrated. In case of classical Ising spins, which can point up or
down with respect to some axis, this leads to an extensive
ground-state degeneracy: for example, in a system of $N$ Ising spins
on a triangular lattice there are $\Omega_N = e^{0.323 N}$
configurations having the same (minimal) energy. On a
three-dimensional pyrochlore lattice of site-sharing tetrahedra, the
(effectively) Ising spins of Dy$_2$Ti$_2$O$_7$, Ho$_2$Ti$_2$O$_7$ and
Ho$_2$Sn$_2$O$_7$ realize~\cite{harris97} fascinating {\em spin ice}
physics~\cite{spinice-bramwell01} where strong local {\em ice rules}
(for any given tetrahedron, two of its spins must point in, and the
other two - out) enforce long-ranged power-law correlations between
spins~\cite{isakov04}, in effect realizing artificial magnetic field
and fractionally charged magnetic monopoles~\cite{monopole08}!

Quantum spins can exploit this extensive degeneracy via
quantum-mechanical coupling between different configurations -- their
wave function can be thought of as a linear superposition of all
degenerate microstates represented by classical patterns of up- and
down-spins.  In the case of strong coupling we may arrive at a {\em
  quantum spin liquid} (QSL)~\cite{pomeranchuk1941,balents2010} state
in which spins never settle in one particular configuration and
continue their exploration forever.  It is clear that such a state
encodes highly nontrivial correlations between different spins when
flipping of one spin induces that of its neighbors so that as a whole
the spin system remains in the lowest-energy manifold. Extensive
experimental~\cite{hiroi2001,shimizu2003,okamoto2007,itou2007,itou2008,olariu2008,yoshida2009,okamoto2009,mendels2010,onoda2010,vachon2011,quilliam2011,yamashita2012,yan2012}
and
theoretical~\cite{anderson1987,wen2002,kitaev2006,meng2010,varney2011,dang2011,white2011,isakov2011,okumura2010,yao2012}
search for materials and models which may realize this intriguing QSL
state constitutes one of the main topics of the quantum spin
physics. Currently there are several intensely researched materials
that hold promise of realizing the elusive spin liquid state. Among
them, we mention two-dimensional spin-1/2 organic triangular antiferromagnets
EtMe$_3$Sb[Pd(dmit)$_2$]$_2$~\cite{itou2007,itou2008,yamashita2012}
and
$\kappa$-(BEDT-TTF)$_2$Cu$_2$(CN)$_3$~\cite{shimizu2003,yamashita2012},
spin-1 material NiGa$_2$S$_4$ \cite{onoda2010},
a series of inorganic quasi-two-dimensional kagom\'e lattice
antiferromagnets: herbersmithite
ZnCu$_3$(OH)$_6$Cl$_2$~\cite{olariu2008,mendels2010}, volborthite
Cu$_3$V$_2$O$_7$(OH)$_2\cdot$2H$_2$O~\cite{hiroi2001,yoshida2009},
and vesignieite
BaCu$_3$V$_2$O$_8$(OH)$_2$~\cite{okamoto2009,quilliam2011}, and a
three-dimensional hyperkagom\'e antiferromagnet
Na$_4$Ir$_3$O$_8$~\cite{okamoto2007}.

Taking the system of frustrated spins to a finite temperature, where
all experiments are done, adds thermal randomness to the picture. At a
finite temperature $T$ even classical spins can explore different
microstates from the lowest-energy manifold. This too leads to a
strongly correlated (although not necessarily phase-coherent, as in
the case of quantum spins at $T=0$) motion of spins which is often
described by a term ``cooperative paramagnet''.  Even if the ground
state of spin system is not a true spin liquid, but instead is one of
the many possible {\em ordered} states, the spins will (thermally) disorder at
sufficiently high temperature, $T \geq T_0$, where $T_0$ stands for
the ordering temperature. In the usual, non-frustrated magnets the
ordering temperature is determined by the exchange interaction energy
$J$ and coordination number of the lattice $z$, $T_0 \sim z S (S+1)
J$. Quite generally, it is of the order of the Curie-Weiss temperature
$\theta_{\rm cw}$ which is easily determined experimentally via
high-temperature behavior of the spin susceptibility $\chi \propto (T
- \theta_{\rm cw})^{-1}$. In antiferromagnets $\theta_{\rm cw}$ is
negative, and, in the absence of frustration, its absolute value sets
the scale at which correlations between spins become pronounced. Thus,
$T_0 \sim |\theta_{\rm cw}|$.  Frustrated magnets are very different
as there $T_0 \ll |\theta_{\rm cw}|$ : despite experiencing strong
interactions with each other, the spins can not ``agree'' on one
particular pattern which would satisfy them all. In the case of a true
spin liquid $T_0 =0$ and the order never arrives.  In the majority of
studied situations the order does take place, $T_0 > 0$, but only at a
temperature much lower than the na\"ive estimate provided by
$|\theta_{\rm cw}|$. This results in a wide temperature interval $T_0
< T < |\theta_{\rm cw}|$ where the spins are strongly interacting but
remain in a disordered cooperative paramagnet state. In fact, the
existence of such temperature (and energy) window represents a
defining feature of the frustrated magnet, as argued by
Ramirez~\cite{ramirez1994} who introduced the frustration parameter $f
= |\theta_{\rm cw}|/T_0$ (it is not uncommon to find a situation with
$f \sim 100$ or greater).

Strong suppression of the ordering temperature can be also the
consequence of close proximity to the quantum critical point,
separating the spin-disordered state (such as a putative QSL) from the
more usual ordered one. It is well established now~\cite{sachdev_book}
that the finite-temperature region above the quantum critical point,
known as the {\em quantum-critical region}, is as informative of the
quantum state of the many-body system as the unreachable $T=0$ ground
state.  The {\em quantum-critical} scaling of, say, the dynamic spin
susceptibility contains information on the spin correlation length,
dynamic exponent $z$ and other critical exponents characterizing spin
systems of different symmetries and dimensions.

Importantly, there is a large number of high-precision
probes---neutron and X-ray scattering, nuclear magnetic resonance,
muon spin rotation, susceptibility, magnetization and specific heat
measurements---which allow us to address various aspects of strange
and conflicting behavior of frustrated quantum magnets experimentally
in a wide range of energies and temperatures.  Given that in many
important cases the correlated spin-liquid region $T_0 < T <
|\theta_{\rm cw}|$ occupies most of the experimentally accessible
temperature interval, unbiased understanding of correlations and
dynamics in this regime becomes the major theoretical task.

Theoretical understanding of frustrated magnetism at finite
temperature is severely limited by the lack of natural small
parameter(s). As a result, possible analytical approaches require one
to study suitably `deformed' models, such as, for example, very
popular and well developed large-component (large-$N$) version of the
Heisenberg spin model on frustrated lattices.  The small parameter is
then provided by $1/N$, expansion in powers of which (about the
$N=\infty$ limit) controls the calculation.  However the physical
limit of $S=1/2$ SU(2) lattice spins corresponds to rather small $N$:
$3$ in the case of $O(N)$ generalization and $1$ in the case of
$Sp(N)$.  Whether or not continuation of the results from
$N=\infty$ to the physical value is reliable remains an open (and case
sensitive) question. Other popular `deformations' include
SU(2)-to-U(1) symmetry reduction(s) and quantum dimer model
approaches. While very insightful and interesting in their own,
applicability of these `modifications' to the original problem is
always an issue.  While analytical approaches are extremely useful in
providing us with qualitative physical insights and understanding, in
frustrated magnets they often fall short of quantitative description
desired by experimentalists.  Numerical approaches are limited as
well. Exact diagonalizations are restricted to small systems (about
$40$ sites at most) due to the exponentially large Hilbert space while
standard quantum Monte Carlo techniques usually suffer from the
infamous `sign problem'. Powerful series expansion methods
often start to diverge in the
most interesting regime $T_0 < T < |\theta_{\rm cw}|$.  Variational
tensor-network type methods are also suffering from finite-size
limitations and are mostly limited to ground state properties.

In this article we combine the most versatile theoretical tool,
Feynman diagrammatics, with the power of Monte Carlo sampling of
complex configuration spaces. The simplest way of arriving at the
diagrammatic technique for spins is to represent them by auxiliary
fermions with imaginary chemical potential.  This trick was introduced
by Popov and Fedotov for spin-1/2 systems in
Ref.~\onlinecite{PopovFedotov1,*PopovFedotov2}.  The ultimate strength
of the diagrammatic approach as compared, e.g.  with high-temperature
or strong coupling expansions, comes from its self-consistent
(skeleton) formulation with automatic summation of certain classes of
graphs up to infinite order.  This lead to better convergence
properties and the possibility of obtaining reliable results in the
strong coupling regime. It turns out that skeleton formulations can be
easily implemented within the sampling protocols leading to the
so-called Bold Diagrammatic Monte Carlo (BDMC)~\cite{bold1} which
obtains physical answers by computing contributions from millions of
graphs and extrapolates them to the infinite diagram order.  Recently,
this method was successfully applied to the normal state of the
Fermi-Hubbard model at moderate interaction strength~\cite{EPL} and
the strongly correlated system of unitary fermions (the so-called
BCS-BEC crossover problem)~\cite{NatureP}. It was, however, never
implemented for models of quantum magnetism, which, at least at the
formal level, can be also viewed as a fermionic system with strong
correlations. Here we present the first attempt to achieve an accurate
theoretical description of the correlated paramagnetic regime within
the BDMC framework.

In what follows in Sec.~\ref{sec:2} we consider a spin-1/2 Heisenberg
type model at finite temperature and its auxiliary fermion version
which admits the standard diagrammatic expansion. In Sec.~\ref{sec:3}
we formulate principles of the BDMC technique and the specific
self-consistent scheme for dealing with skeleton diagrams based on
fully dressed lines. We proceed with detailed description of the worm
algorithm for efficient sampling of the resulting configuration space
(updates, counters, and data processing) in Sec.~\ref{sec:4}.  Our
results for the triangular lattice antiferromagnet are presented,
discussed, and compared to the best available finite temperature simulations based on
numerical linked-cluster (NLC) expansions~\cite{Rigol} and extrapolations of high-temperature series
~\cite{singh05} in Sec.~\ref{sec:5}. We conclude with broader implications of this work
and further developments in Sec.~\ref{sec:7}.

\section{Model and its `bold-line' diagrammatic expansion}
\label{sec:2}

Consider the standard Heisenberg model
\begin{equation} \label{HM}
  H = \sum_{i,j} J_{ij} \, {\vec S}_i \cdot {\vec S}_j  \, ,
\end{equation}
where ${\vec S}$ are quantum spin-1/2 operators. The dimension of
space, lattice geometry, and interaction range are assumed to be
arbitrary. Now, the idea is to replace spin degrees of freedom with
fermionic ones:
\begin{equation} \label{SF}
  {\vec S}_i \to \frac{1}{2}\sum_{\alpha, \beta} f_{i \alpha}^{\dagger }
  \vec{\sigma }_{\alpha \beta} f_{i \beta}^{\,}   \, .
\end{equation}
Here $f_{i \beta}$ is the second quantized operator annihilating a
fermion with spin projection $\beta =\pm 1$ on site $i$, and
$\vec{\sigma }$ are the Pauli matrices. As a result, we convert
coupling between spins into the standard two-body interaction term
$J_{ij,\alpha \beta \gamma \delta} \, f_{i \alpha }^{\dagger }f_{i
  \beta}^{\,} f_{j \gamma}^{\dagger }f_{j \delta}^{\,}$ with matrix
element $J_{ij, \alpha \beta \gamma \delta}=(1/4)J_{ij} \vec{\sigma
}_{\alpha \beta} \cdot \vec{\sigma }_{\gamma \delta}$.  By doing so we
also increase the Hilbert space on every site from 2 to 4 by adding
non-physical states with zero and two fermions.

The benefit of the auxiliary fermion representation does not require
explanation: once the spin model is mapped onto a familiar problem of
interacting fermions one can employ numerous diagrammatic tricks to
solve it. However, this raises an issue of eliminating contributions
from unphysical states to the answer in a manner consistent with the
diagrammatic technique.  Remarkably, there is a very simple way to
achieve the goal by adding the chemical potential term to the
fermionic Hamiltonian~\cite{PopovFedotov1,*PopovFedotov2} (see also
Ref.~\onlinecite{Fermionization})
\begin{equation} \label{HF}
  H_F = \sum_{ij,\alpha \beta \gamma \delta }J_{ij, \alpha \beta \gamma \delta } \,
  f_{i \alpha}^{\dagger }f_{i \beta}^{\,} f_{j \gamma}^{\dagger }f_{j \delta}^{\,} - \mu \sum_{i} (n_{i} -1)
  \, ,
\end{equation}
with complex $\mu = -i\pi T/2$ and $n_{i} =\sum_\alpha f_{i
  \alpha}^{\dagger }f_{i \alpha}^{\,}$.  The added term commutes with
the original Hamiltonian and has no effect on properties of the
physical subspace $\{ n_i=1 \}$ whatsoever. Moreover, the grand
canonical partition functions and spin-spin correlation functions of
the original spin model and its fermionic version are also identical
because (i) physical and non-physical sites decouple in the trace and
(ii) the trace over non-physical states yields identical zero on every
site.  As a result, we arrive at a rather standard Hamiltonian for
fermions interacting through two-body terms. Complex value of the
chemical potential is essentially a zero price to pay for the luxury
of having the diagrammatic technique.

Formally, the entire setup is similar to the fractional quantum Hall
effect system because the non-interacting part of Eq.~\eqref{HF}
describes particles with zero dispersion relation, i.e. we start
building the solution from the degenerate manifold. There is one
important difference though: in conventional fermionic systems both
the Green's function and the spin-spin (or density-density)
correlation function contain direct physical information about the
system.  In spin systems the Green's function is rather an auxiliary
object which always remains localized on a single site. This does not
imply any pathological behavior yet since the physical degrees of
freedom are spins, i.e. bilinear combinations of fermionic operators.
Correspondingly, the main object of interest is not the Green's
function of the system but the spin-spin correlation function, or
magnetic susceptibility:
\begin{equation}
  \chi (i,j,\tau ) = \langle T_{\tau}  S^{z}_i(0 ) S^{z}_j(\tau ) \rangle =\frac{1}{3} \langle T_{\tau}  {\vec S}_i(0 ) \cdot {\vec
    S}_j(\tau ) \rangle  \, ,
 \label{chi}
\end{equation}
where $T_{\tau}$ stands for the imaginary time ordering operator.

To simplify the presentation below we consider lattices with one atom
per unit cell and make use of the lattice translation invariance;
otherwise one would need to keep an index enumerating different sites
in the unit cell.  For the same reason we do not place the system into
the external magnetic field to preserve the symmetry between up- and
down-spins; in the presence of the magnetic field $H_0$ one has to add spin-dependent real part
to the chemical potential $\mu \to \mu_\alpha = -i\pi T/2 - \alpha
H_0$ and take proper care of the spin index.

\begin{figure}[htbp]
\includegraphics[angle=0,width=1.\columnwidth]{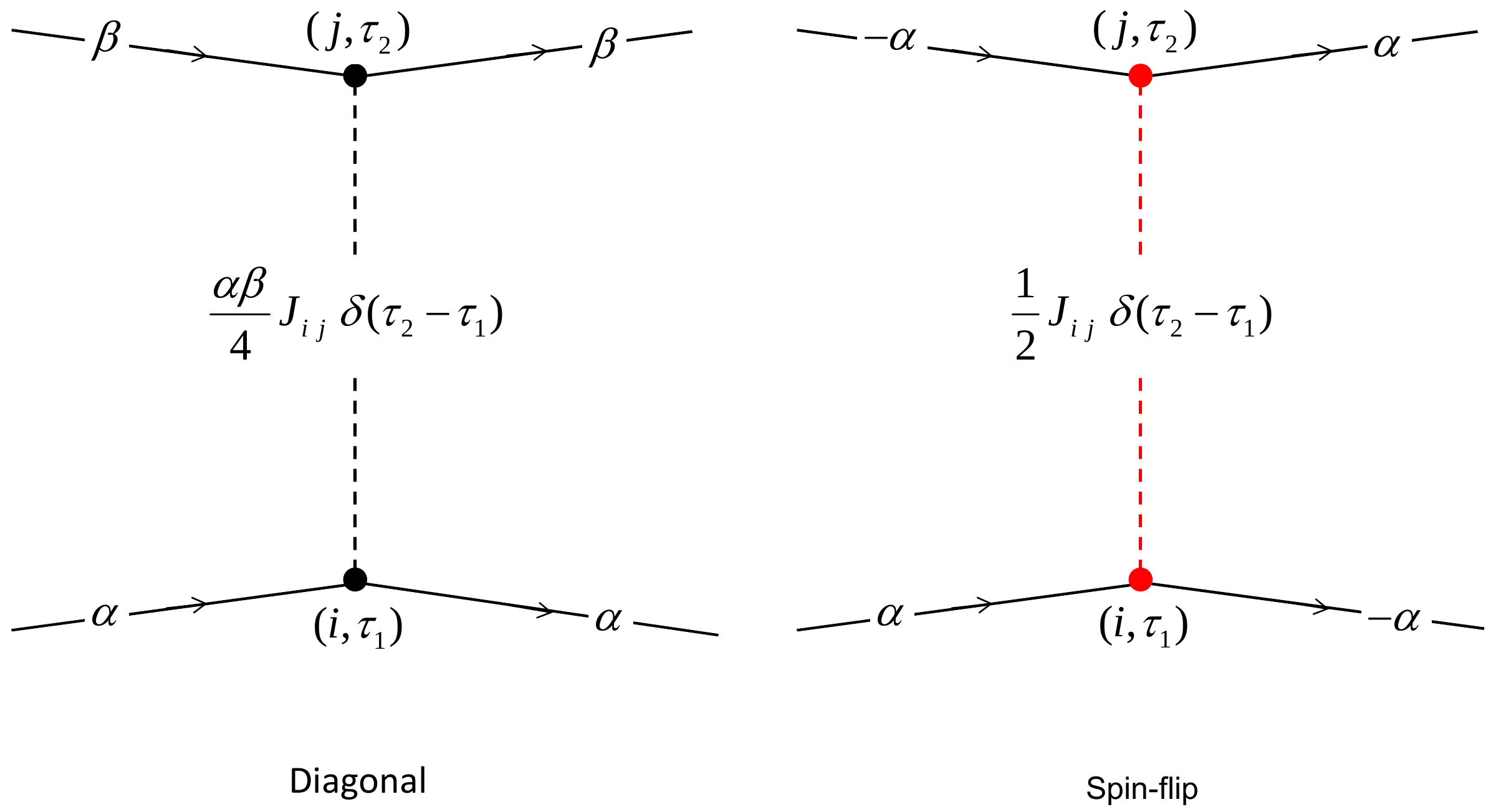}
\caption{\label{fig:1}
  (Color online) Graphic representation of allowed interaction
  processes for the Heisenberg model.
}
\end{figure}

The diagrammatic technique itself for Eq.~\eqref{HF} is absolutely
standard~\cite{FW}.  The perturbative diagrams are expressed in terms
of the non-interacting, or 'bare', Green's functions (particle
propagators), $G^{(0)}_{\alpha\beta} (i, \tau ) \equiv \delta_{i,0}
\delta_{\alpha , \beta} G^{(0)} (\tau )$, and two-body interaction
lines $J_{i-j, \alpha \beta \gamma \delta}$.  The non-interacting
Green's function on a lattice in the imaginary time representation
reads (below the Boltzmann constant $k_B=1$)
\begin{equation} \label{G0}
  G^{(0)} (\tau  > 0 ) = - \, \frac{e^{\mu \tau}}{1+\exp (\mu/T)} =
  \frac{e^{\mu \tau}}{i-1} \, ,
\end{equation}
with conventional anti-periodic boundary conditions $G^{(0)} (\tau <
0) = - G^{(0)}(1/T + \tau )$.  In what follows we completely suppress
the site index for purely local quantities.  For the Heisenberg model
the dependence of the interaction line on spin indexes is rather
simple
\begin{eqnarray} \label{J0}
  J_{i-j, \alpha \beta \gamma \delta} & = & \frac{J_{i-j}}{4} \,
  \hat{M}_{\alpha \beta \gamma \delta} \,, \nonumber \\
  \hat{M}_{\alpha \beta \gamma \delta} & = &
  \alpha \gamma  \, \delta_{\alpha ,\beta} \delta_{\gamma  ,\delta} +
  2 \, \delta_{\delta ,\alpha} \delta_{\gamma ,-\alpha} \delta_{\beta ,-\alpha}
  \, ,
\end{eqnarray}
The first term describes diagonal coupling between the spin densities
on sites $i$ and $j$, while the second spin-flip term exchanges spin
values, see Fig.~\ref{fig:1}.  The magnitude of the spin-flip process
is fixed by the SU(2) symmetry of the problem. Even when interactions
are screened by many-body effects, see below, the retarded interaction
has exactly the same dependence on spin indexes.  An arbitrary
perturbative diagram of order $n$ contributing to, say, the
free-energy is obtained by (i) placing graphical elements depicted in
Fig.~\ref{fig:1} with some space/time variables $i$, $j$, and $\tau
\in (0,1/T)$ and connecting incoming and outgoing propagators with the
same spin and site index to each other in such a way that all points
in the resulting graph are connected by some path.
%

\section{Bold Diagrammatic Monte Carlo scheme}
\label{sec:3}

The unique feature of diagrammatic expansions for propagators is that
there are no numerical coefficients in the diagram weight depending on
the diagram order or structure (this is not true for other well-known
series such as virial, high-temperature, linked-cluster,
strong-coupling, etc. expansions). This leads to the diagrammatic
technique when certain infinite sets of diagrams, e.g. in the form of
geometric series, are easily dealt with by algebraic means or reduced
to self-consistently defined integral equations.

\begin{figure}[htbp]
\includegraphics[angle=0,width=1.\columnwidth]{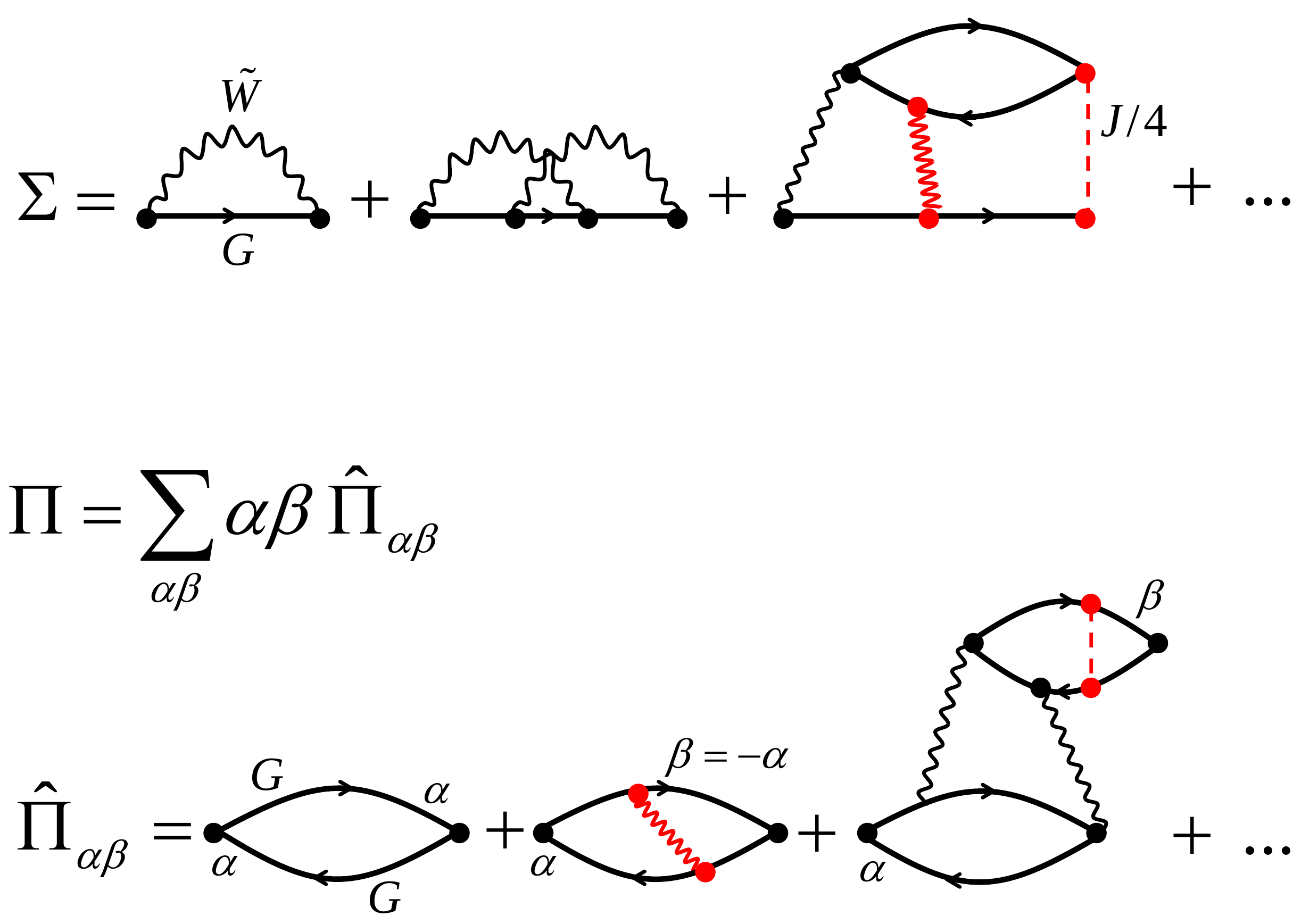}
\caption{\label{fig:2}
  (Color online) Typical low-order diagrams contributing to the
  particle self-energy and polarization operator within the
  $G^2W$-skeleton scheme (red color denotes spin-exchange
  coupling). Diagrams for $\Sigma$ and $\Pi$, in their turn, are used
  to calculate fully dressed $G$ and $W$ functions, see
  Eq.~\eqref{dysonGW}.
}
\end{figure}

In the skeleton technique, the diagrams are classified according to
some rule which eliminates the need for computing repeated blocks of
diagrams. In the simplest scheme which is used in this paper, one
identifies the proper self-energy blocks which consist of diagrams
where all vertexes (points where two particle propagators and the
interaction line meet) remain connected by some path when one removes
any two lines of the same kind: two propagator lines with the same
spin index or two interaction lines. We will refer to this set of
diagrams as {\it irreducible}.  The omitted diagrams are fully
accounted for by replacing bare propagators and interaction lines in
the proper self-energy blocks with exact propagators, $G(\tau)$, and
screened interactions, $W(r,\tau )\hat{M}$. The resulting formulation
is self-consistent and highly non-linear since $G$ and $W$ depend on
proper self-energies through the Dyson type equations. If $\Sigma $ is
the proper self-energy for the particle propagator, and $\Pi $ is an
analogous quantity for the interaction line (better known as
polarization operator) then (in Fourier representation $(r,\tau ) \to
(q,\omega_m)$ for space-time variables)
\begin{eqnarray}
 &&G(m) = \frac{G^{(0)}(m) }{1 - G^{(0)}(m) \Sigma (m) }  \, , \nonumber \\
 &&W (q,m) = \frac{J(q)}{4 - J(q) \Pi (q,m)} \equiv \frac{J(q)}{4} + \tilde{W}(q,m) \, .
\label{dysonGW}
\end{eqnarray}
where $J(q)=\sum_r e^{iqr} J(r)$. Due to fermionic/bosonic nature of
propagators $G$/$W$ we have different definitions of the Matsubara
frequency here, $\omega_m=2\pi T(m+1/2)$ for $(G,\Sigma )$ and
$\omega_m=2\pi T m$ for $(W,\Pi )$.  Note also that we split the $W$
function into two parts by separating out the original coupling.  This
is done for technical reasons explained below; here we simply point
out that in the imaginary time domain this is equivalent to paying
special attention to the $\delta$-functional contribution, $W(q,\tau ) =
\tfrac14 J(q)\, \delta (\tau) + \tilde{W}(q,\tau )$.  Correspondingly,
for graphical representation of the diagrams we use wavy lines for
$\tilde{W}$ and dashed vertical lines for bare coupling.
The corresponding (rather standard in many-body theory) $G^2W$-skeleton
scheme is illustrated in Fig.~\ref{fig:2}.
The magnetic susceptibility Eq.~\eqref{chi} is directly related to
the polarization operator appearing in Eq.~\eqref{dysonGW} and Fig.~\ref{fig:2}
\begin{equation}
\chi(q,n) =  \frac{\Pi (q,n)}{4 - J(q) \Pi (q,n)} \, .
\label{dysonGWchi}
\end{equation}
%

To make connection with general rules of diagrammatic
MC~\cite{polaron98,polaron2000} we formalize the problem at hand as
computing quantity $Q(y,s)$ (where $y$ stands collectively for space,
imaginary time, and spin variables, while $s=1,2$ labels the proper
self-energy and the polarization operator sectors) from the series of
multidimensional sums/integrals
\begin{equation}
Q(y,s) = \sum_{n \xi} \int dx_1 ...
dx_n dY\;
{\cal D} (n,\xi , \{ x_i \} ; Y,s ) \delta(y-Y) \; .
\label{D}
\end{equation}
Here $n=0,1, \dots \infty$ is the diagram order, $\xi$ labels
different terms of the same order, $x_i$ are internal
integration/summation variables, and ${\cal D}$ is the diagram
contribution to the answer.  Formally, one can think of the set of
skeleton diagrams for the free-energy of the system with one line
marked as 'dummy'. When the dummy line is removed from the graph the
rest is interpreted as a diagram for $\Sigma (y)$ (in this case $s=1$
and the dummy line is the particle propagator) or $\Pi (y)$ (in this
case $s=2$ and the dummy line is the $\tilde{W}$ one). The last rule
follows from the fact that $\Pi (y)$ is a continuous function of time.
In the BDMC approach one interprets Eq.~\eqref{D} as averaging of
$e^{i \arg({\cal D} ) }\delta(y-Y)$ over the configuration space $\nu =
(n,\xi , x_1, \dots , x_n ; Y,s )$ with probability density
proportional to $|{\cal D} |$.

The skeleton formulation does {\it not} cause any fundamental problem
for Monte Carlo methods and is easy to implement for diagrams of
arbitrary order.  Essentially, at any stage in the calculation both
$G$ and $W$ are considered to be known functions (the calculation may
start with $G=G^{(0)}$ and $W=J/4$) while Eq.~\eqref{dysonGW} is used
from time to time to improve one's knowledge about $G$ and $W$ using
accumulated statistics for $\Sigma$ and $\Pi$ and fast Fourier
transform algorithms.  Moreover, we have shown~\cite{bold1} that BDMC
methods are more stable and have better convergence properties than
conventional iterations.  One immediately recognizes that in the
skeleton-type formulation (i) the number of diagrams to be sampled in
a given order is dramatically reduced, (ii) the convergence of the
skeleton series is likely to be different than that of the bare
series, (iii) non-analytic and non-perturbative behavior might emerge
even from a finite number of terms due to highly non-linear
self-consistent formulation (we refer here to the famous mean-field
BCS solution).

The other crucial advantage of BDMC over more conventional MC methods
simulating finite clusters of spins is that it deals directly with the
thermodynamic limit of the system.  In practice, in dimension $d>1$
the error bars often become too large before a reliable extrapolation
to the thermodynamic limit can be done.  In this sense, BDMC is not
subject to the infamous sign-problem which is understood as
exponential scaling of computational complexity with the space-time
volume of the physical system. Feynman diagrams do alternate in sign
and contributions from high-order diagrams cancel each other to near
zero. However, this behavior is better characterized as a
'sign-blessing', not a sign-problem, because it is crucial for
convergence properties.
With the number of graphs growing factorially
with their order the only possibility for obtaining series with finite
convergence radius is to have sign-alternating terms such that
high-order diagrams cancel each other (the sign-blessing phenomenon).
For series with finite
convergence radius there are numerous unbiased re-summation techniques
which allow one to determine the answer well outside of the
convergence radius provided enough terms in the series are
known. Relatively small configuration space for skeleton diagrams
allows one to establish if sign-blessing takes place in a given model
and to obtain accurate results for diagram orders as large as 7-10
(depending on the model).

It should be noted that in recent years the Popov-Fedotov trick has
become popular within the functional renormalization group (PFFRG)
framework.  It has been applied to frustrated $J_1 - J_2$ model on
square lattice~\cite{wolfle2010}, planar $J_1 - J_2 - J_3$
antiferromagnet~\cite{reuther2011}, spatially anisotropic triangular
antiferromagnet~\cite{thomale2011}, and honeycomb lattice
antiferromagnet with competing interactions~\cite{abanin2011}. These
studies also attempt at attacking the problem using Feynman
diagrammatic series but are radically different in the technical
implementation.  While PFFRG is also based on sums of subsets of
selected diagrams to infinite order, it does not offer a convenient
way to check for convergence of final results as more and more
diagrams are retained. This is the most severe drawback of PFFRG;
after all the major problem with existing theories is reliable
estimate of the accuracy. The PFFRG spectral functions are very broad
in energy~\cite{wolfle2010} and appear to underestimate ordering
fluctuations. This also leads to significant rounding of
susceptibility peaks and makes identification of different phases
difficult. In addition, these studies are typically focused on the
zero-temperature phase diagram, not the finite-temperature cooperative paramagnet state.

\section{Normalization and worm-algorithm updates}
\label{sec:4}

To simplify notations let us write the diagrammatic contribution from
the configuration space point $\nu$ as ${\cal D}_{\nu} =
e^{i\varphi_{\nu}} D_{\nu}$ and call the non-negative function
$D_{\nu}$ the configuration 'weight'.  Within the $G^2W$-skeleton
scheme $D_{\nu}$ is given by the modulus of a product which runs over
all lines
\begin{equation}
D_{\nu} = \vert \prod_{\rm lines} f_{\rm line}(\nu ) \vert  \; ,
\label{D0}
\end{equation}
where $f_{\rm line}(\nu )$ stands for a collection of functions describing various
lines in the diagram.
At this point we notice that equation
\begin{equation}
Q(y,s) = \sum_{\nu} e^{\varphi_{\nu}}  D_\nu  \delta(y-Y)\; .
\label{D1}
\end{equation}
can be always interpreted as averaging over the probability density distribution
$P_{\nu s} = D_{\nu}/C_s$, where $C_s$ is the normalization factor,
and thus sampled by MC methods:
\begin{equation}
Q(y,s) =  C_s \sum_{\nu} e^{i\varphi_{\nu}} \delta(y-Y) P_{\nu s} \longrightarrow
C_s \sum_{\nu}^{\MC} e^{i\varphi_{\nu}} \delta(y-Y)  \;.
\label{D2}
\end{equation}
In the last transformation we replace the full sum over the
configuration space with the stochastic sum over configurations which
are generated from the probability density $P_{\nu s}$. This is, of
course, nothing but the standard MC approach to deal with complex
multi-dimensional spaces. We stress here that {\it all} configuration
parameters are sampled stochastically, including the diagram order and
its structure, making Diagrammatic MC radically different from calculations
which first create a list of all diagrams up to some high-order and
then evaluate them one-by-one (often with the use of MC methods for
doing the integrals).

\subsection{Normalization}

The normalization constant $C$ can be determined in a number of ways: \\
{\bf (i)} using known behavior of $Q(y,s)$ in some limiting case, for
example
$Q(y \to y_0,s) \to Q_0(s)$,  \\
{\bf (ii)} through the exact sum rule, $\int dy Q(y,s) = R_s$,
if available, or, more generically, \\
{\bf (iii)} by measuring the ratio between the contributions of all
diagrams in Eq.~\eqref{D} and diagrams which are known either
analytically or numerically with high accuracy.  Indeed, imagine that
one or several diagrams, say the lowest order ones, are known and
their integrated contribution to the answer is $Q_N(s) =\int dy
Q_N(y,s)$. Let ${\cal N}_s$ be their configuration space.  Then the
ratio $Q(y,s)/Q_N(s)$ can be measured in the MC simulation as
\begin{equation}
  \left[ \int dy Q(y,s) \right] /Q_N(s) = (\sum_{\nu}^{\MC} e^{i\varphi_{\nu}})/(
  \sum_{\nu}^{\MC} \delta_{\nu \in {\cal N}_s} \; e^{i\varphi_{\nu}}  ) \;.
\label{N0}
\end{equation}
This leads to
\begin{equation}
C_s = \frac{Q_N(s)}{Z_s}\;, \qquad \qquad
Z_s = \sum_{\nu}^{\MC}\delta_{\nu \in {\cal N}_s}\;  e^{i\varphi_{\nu}}  \;.
\label{N1}
\end{equation}

The diagrams used for normalization are not necessarily the physical
ones, i.e.  they can be artificially ``designed'' to have simple
analytic structure and added as a special sector to the configuration
space $\{ \nu \}$, see Ref.~\onlinecite{bold1}. In the latter case,
the diagrams contributing to $Q(y,s)$ and $Z_s$ are mutually exclusive
and physical contributions have to be filtered by $(1-\delta _{\nu \in
  {\cal N}_s})$.

In the present study we use the modulus of the Hartree diagram to
normalize statistics for $\Sigma$, and the modulus of the lowest order
GG-diagram, see the first term in Fig.~\ref{fig:2}, to normalize
statistics for $\Pi$:
\begin{eqnarray}
Q_N(1)&=&\Sigma_N =  \sum_r \sum_{\alpha } \vert \frac{J(r)}{4} \vert \, \vert G(\tau = -0) \vert   \;, \nonumber \\
Q_N(2)&=&\Pi_N    =  \sum_{\alpha } \int_0^{1/T} d\tau  \vert G(\tau ) G(-\tau ) \vert    \;.
\label{N2}
\end{eqnarray}
Even though in the self-consistent scheme one does not know the
$G(\tau )$-function analytically (it is tabulated numerically) it
takes no time to compute the normalization factor $\Pi_N$ with high
accuracy. We take the modulus of the Hartree diagrams because in the
absence of external magnetic field the spin-up and spin-down
contribution exactly cancel each other. Correspondingly,
\begin{eqnarray}
Z_1 &=&  \sum_{\nu}^{\MC} \delta_{\nu , ({\rm Hartree},s=1)}  \;, \nonumber \\
Z_2 &=&  \sum_{\nu}^{\MC}\delta_{\nu ,(n=1,s=2)}                 \;.
\label{N3}
\end{eqnarray}

\subsection{Worm-algorithm trick}

We now proceed with the description of updates which ensure that all
points in the configuration space are sampled from the probability
density $P_{\nu s}$. Among many possibilities we seek a scheme which
involves the smallest number of lines in a single update and does not
require global analysis of the diagram structure. Such updates are
called "local". Typically, they are much more flexible in design, are
easier to implement, and lead to more efficient codes by having large
acceptance ratios. The reader not interested in algorithmic details
may proceed directly to the next Section.

An easy way to ensure that the diagram is irreducible is to check that
no two lines in the graph have the same momentum---by momentum
conservation laws the isolated self-energy blocks cannot change the line
momentum. By creating a hash table where momenta of all lines are
registered according to their values one can readily verify that
momenta of updated lines are not repeated in the graph without looking
at the graph topology or addressing all other lines. This simple tool
solves the problem of performing local updates within the irreducible
set. It can be applied even if diagrams are sampled in the real-space
representation, as is done in this article, by attaching an auxiliary
momentum variable to each line and satisfying the momentum
conservation law at each vertex. In this case, the sole purpose of
introducing auxiliary momenta is an efficient monitoring of the
diagram topology and the ${\cal D}_\nu$ value is independent of them.

Since the trick with auxiliary momenta is based on momentum
conservation laws one is necessarily limited to either (i) performing
updates on closed loops, i.e. partially abandoning the idea of local
updates, or (ii) extending the configuration space to include diagrams
which violate these conservation laws. The second strategy is the
essence of the worm algorithm. One more reason for using the worm
algorithm approach is the spin projection conservation law in the
interaction process, see Eq.~\eqref{J0} and Fig.~\ref{fig:1}. It also
requires that updates are performed only on closed loops of
interaction lines and propagators, unless one admits diagrams which
violate the corresponding conservation law. It turns out that a
straightforward extension of the worm algorithm introduced in
Refs.~\onlinecite{Kris08} and \onlinecite{EPL} allows one to go around
both hurdles.

\begin{figure}[htbp]
\includegraphics[angle=0,width=1.\columnwidth]{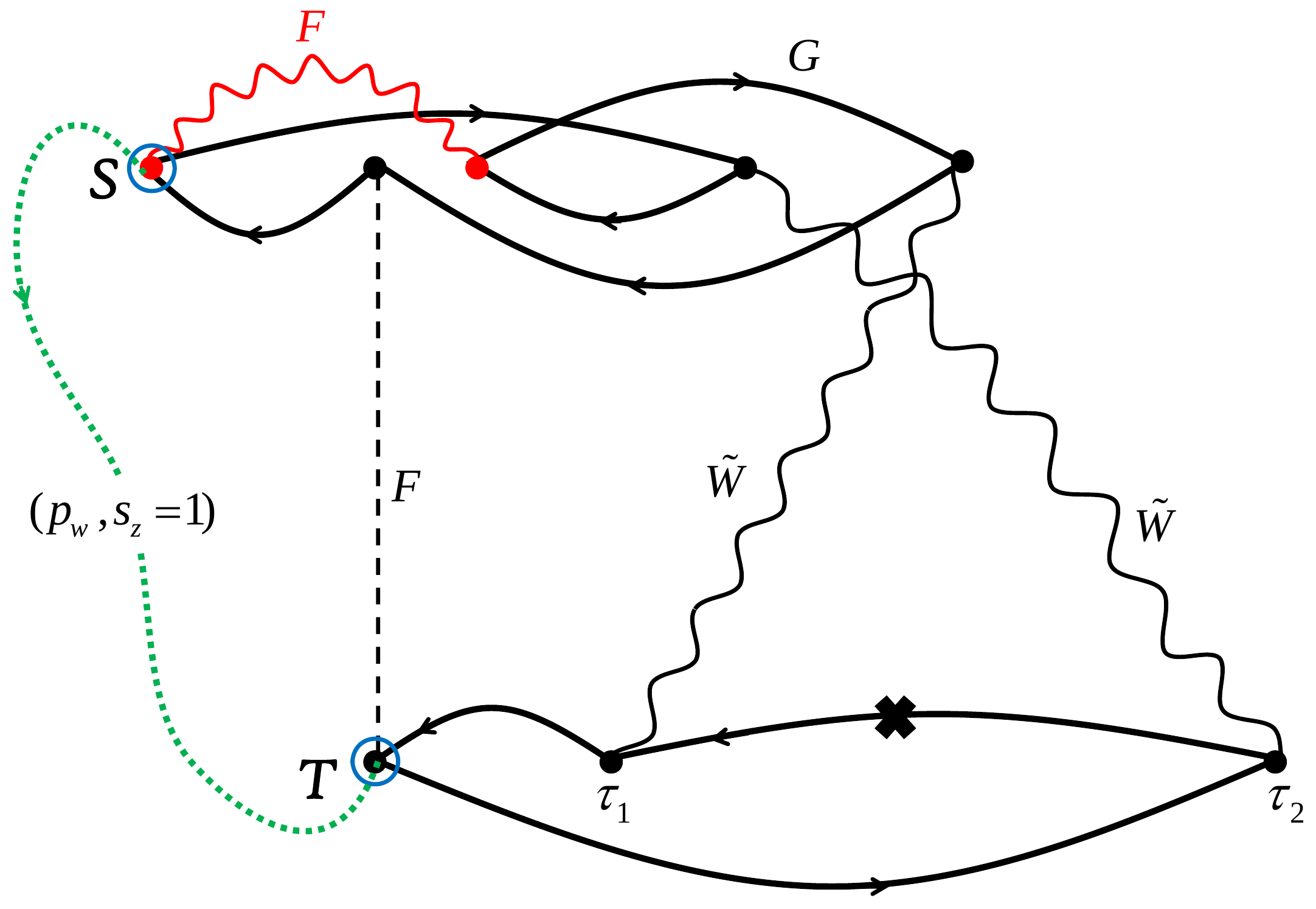}
\caption{\label{fig:3}
  (Color online) Non-physical diagram with two special vertexes
  (worms) ${\cal S}$ and ${\cal T}$ (marked by blue circles). Momentum
  and spin conservation laws would be satisfied at all vertexes and
  interaction lines if one considers ${\cal S}$/${\cal T}$ as a
  drain/source of momentum $p_w$ and spin projection $1$; the same
  rule is recovered by imagining a line (green dotted) which carries
  $(p_w, s_z=1)$ from ${\cal S}$ to ${\cal T}$. If special vertexes
  were not present in the diagram, then this graph would contribute to
  $\Sigma (\tau_1-\tau_2)$ after removing the dummy propagator line
  marked by the cross (if cross marks the dummy $\tilde{W}$-line then
  the diagram is contributing to $\Pi$).
  }
\end{figure}

\begin{figure}[htbp]
\includegraphics[angle=0,width=1\columnwidth]{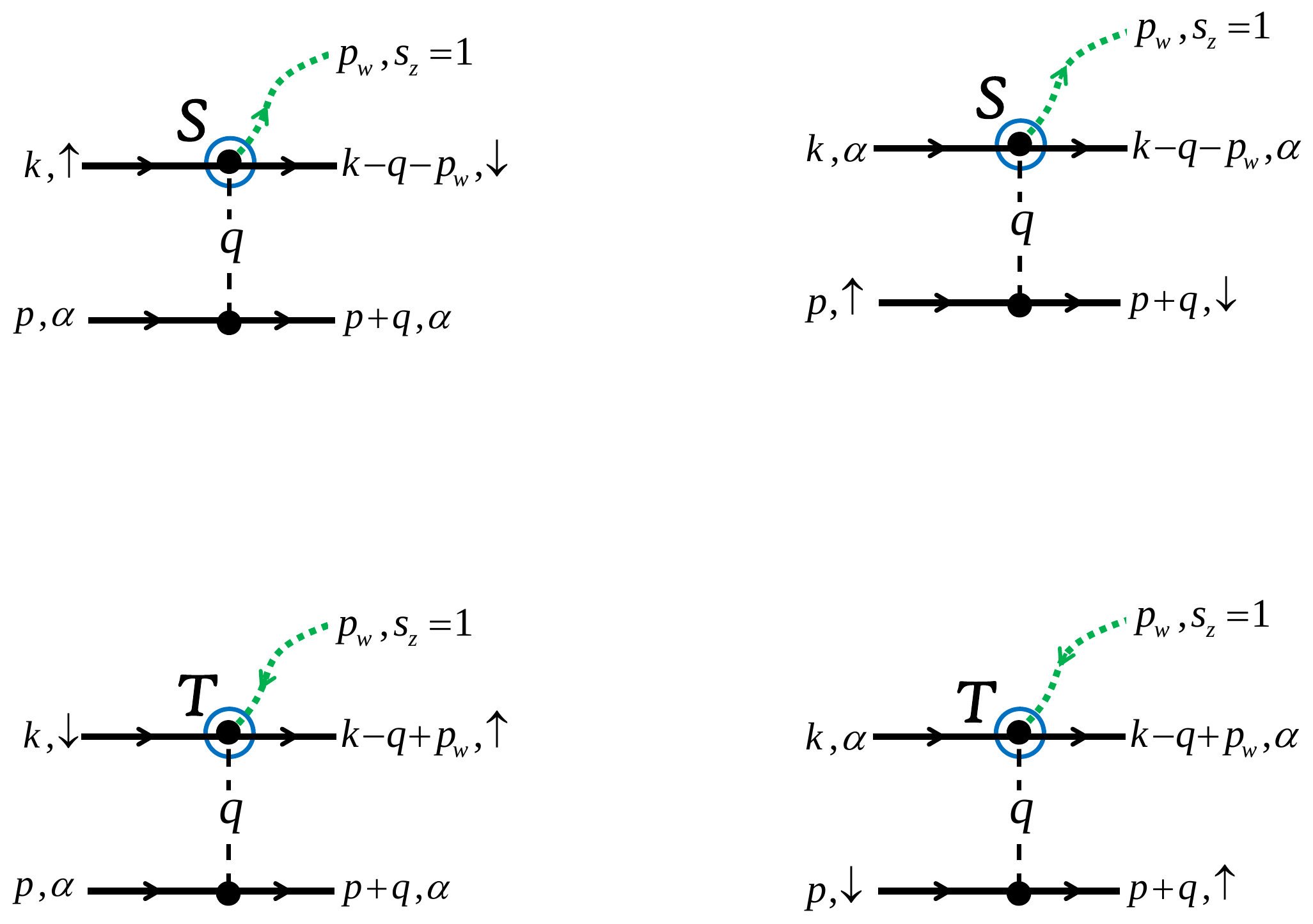}
\caption{\label{fig:4}
  (Color online) Detailed structure of ${\cal S}$ and ${\cal T}$
  vertexes. Note, that the interaction line containing a worm on one
  of its ends is also unphysical and the spin conservation law applies
  to a pair of vertexes.
}
\end{figure}

The additional (unphysical) diagrams have the following
structure. There are two special vertexes, or 'worms', ${\cal S}$ and
${\cal T}$, where momentum and spin conservation laws are violated,
see Fig.~\ref{fig:3}.  In a given graph, any two vertexes can be
special if they are not connected by the interaction
line. Conservation laws become satisfied if one imagines a special
line connecting ${\cal S}$ to ${\cal T}$.
This line is not associated with any propagator and its mission is to transfer momentum
$p_w$ and spin projection $1$ (doted line in Fig.~\ref{fig:3});
otherwise special vertexes represent a drain and a source of momentum
$p_w$ and spin projection $1$, see Figs.~\ref{fig:3} and \ref{fig:4}.
It has to be realized that the spin conservation law is formulated for
a pair of vertexes connected by the interaction line, i.e. an
interaction line containing a worm on one of its ends is unphysical
and can be described by any suitable function $F(r, \tau)$.
%
%

The worm algorithm idea is based on the observation that updates
performed with the use of special vertexes can always be made local,
including non-trivial changes in the diagram topology and order as
well as transformations replacing diagonal interaction lines with
spin-flip ones. Even though unphysical diagrams are frequently
encountered in the simulation process they are excluded from the
statistics of $\Sigma$ and $\Pi$ which is accumulated only on the
physical set of diagrams.

\subsection{Updates}

Below we describe the simplest ergodic updating scheme. With trivial
modifications and additional filters for proposals which would be
rejected because they are incompatible with the allowed configuration
space it can be made more efficient. This, however, would overwhelm
the presentation with minor programming details and we choose not to
discuss them here.  For example, below we will use the following
algorithmic rules for dummy lines used to identify diagrams as
$\Sigma$- or $\Pi$-type: (i) the dummy interaction line cannot be
removed or created in any update, (ii) if the dummy propagator
originating from vertex ${\cal A}$ is modified by adding/removing an
intermediate vertex ${\cal C}$ then ${\cal A}$ always remains the
originating vertex of the new dummy line. These rules can be easily
modified to avoid fast rejections of updates in certain cases at the
expense of using additional random numbers to deal with available
choices and taking care of them in acceptance ratios. Alternatively,
the notion of the dummy line can be avoided altogether by designing an
improved estimator based on free-energy diagrams.

\begin{figure}[htbp]
\includegraphics[angle=0,width=1.\columnwidth]{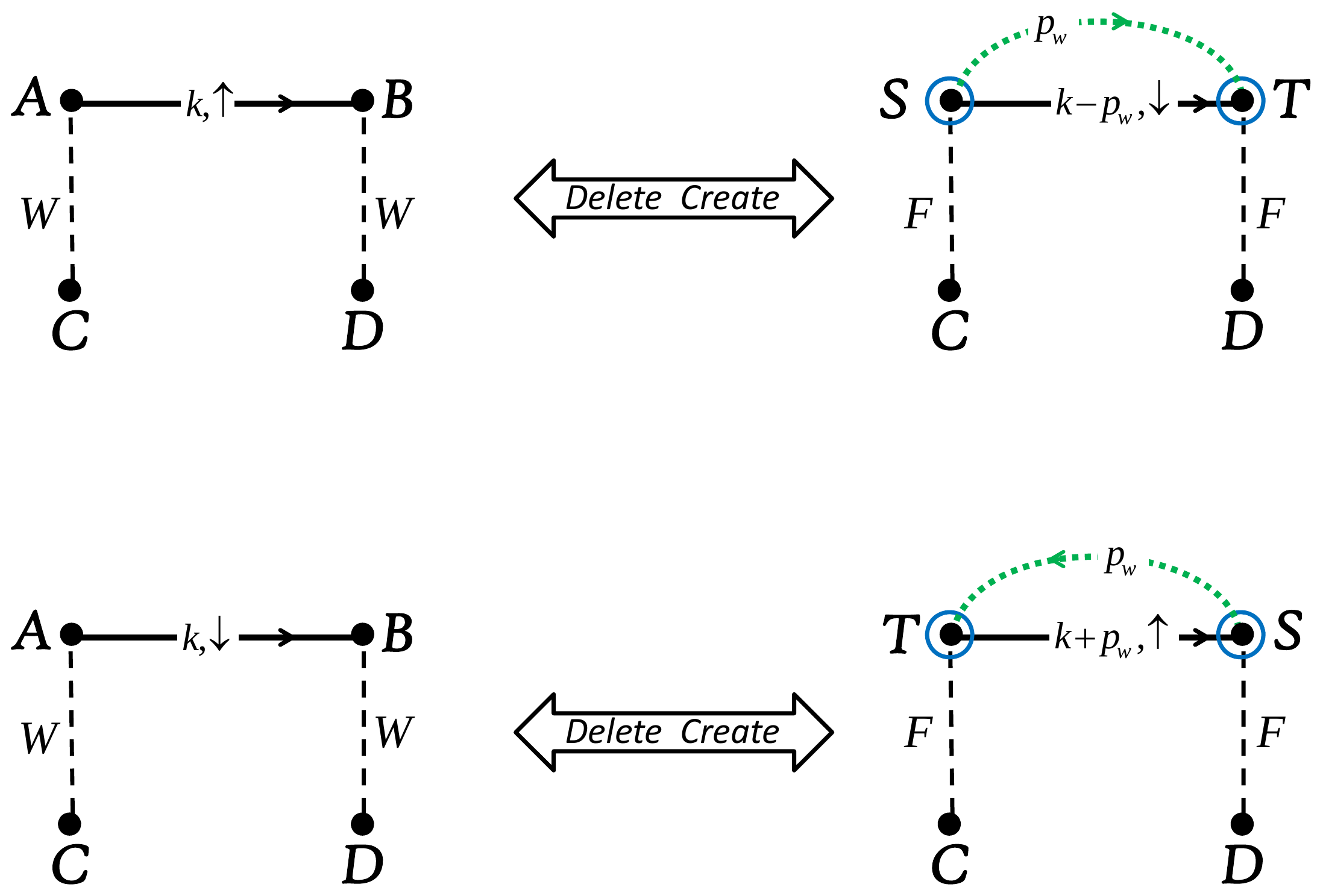}
\caption{\label{fig:5}
  Two cases for {\it Create} and {\it Delete} updates which
  insert/remove a pair of worms at the ends of the particle
  propagator.
}
\end{figure}

\paragraph*{\underline{Create - Delete}}
The pair of complementary updates {\it Create} - {\it Delete} switches
between physical and unphysical sectors by inserting/removing a pair
of special vertexes connected by the particle propagator.  It is not
allowed to have ${\cal S}$ = ${\cal T}$ or to have ${\cal S}$ and
${\cal T}$ being connected by the interaction line, see an
illustration in Fig.~\ref{fig:5}.  In {\it Create}, the particle
propagator for update and the type of special vertex to be placed at
its origin (${\cal A} \to {\cal S}$ or ${\cal A}\to {\cal T}$) are
selected at random.  One has to verify that flipping the propagator
spin is consistent with the worm rules or reject the proposal.  The
missing momentum $p_w$ at the worm vertex is selected at random.  In {\it Delete}, one
selects a worm at random, checks that the outgoing particle propagator
arrives at the other worm, and proposes to remove worms from the
diagram.  The corresponding acceptance ratios are given by
\begin{equation}
R_{\rm Create}=\frac{D_{\nu '}}{D_{\nu}} \, \frac{u_2}{u_1}\, 2n\,   \;, \;
R_{\rm Delete}=\frac{D_{\nu '}}{D_{\nu}} \, \frac{u_1}{u_2}\, \frac{1}{2n} \;,
\label{U12}
\end{equation}
where $n$ is the diagram order and $\nu$ and $\nu '$ are configuration space points before and
after the update and $D_{\nu}$ is the product in Eq.~\eqref{D0}. In the
present case only one propagator and two interaction lines are
affected:
\begin{equation}
  \frac{D_{\nu '}}{D_{\nu}} = \left\vert \frac{G_{AB}(\alpha ) F_{SC}
      F_{TD}}{G_{AB}(-\alpha ) W_{AC} W_{BD}}
  \right\vert \, ,
  \label{DD}
\end{equation}
in {\it Create} and similarly, $D_{\nu '}/D_{\nu} = |GWW/GFF|$, in
{\it Delete} with appropriate arguments for all functions
involved. [If the line is labeled as $W$ it can be either of $J$ or
$\tilde{W}$ type.] In what follows we will stop mentioning which
vertexes determine function parameters since these can be easily
recovered from the figures.  Finally, assuming that the protocol of
deciding which update should be implemented next is random and based
on assigning each update some probability, $u_i$, we mention the ratio
of probabilities in the acceptance ratio ($u_1$ for {\it Create} and
$u_2$ for {\it Delete}).
%

\begin{figure}[htbp]
\includegraphics[angle=0,width=1.\columnwidth]{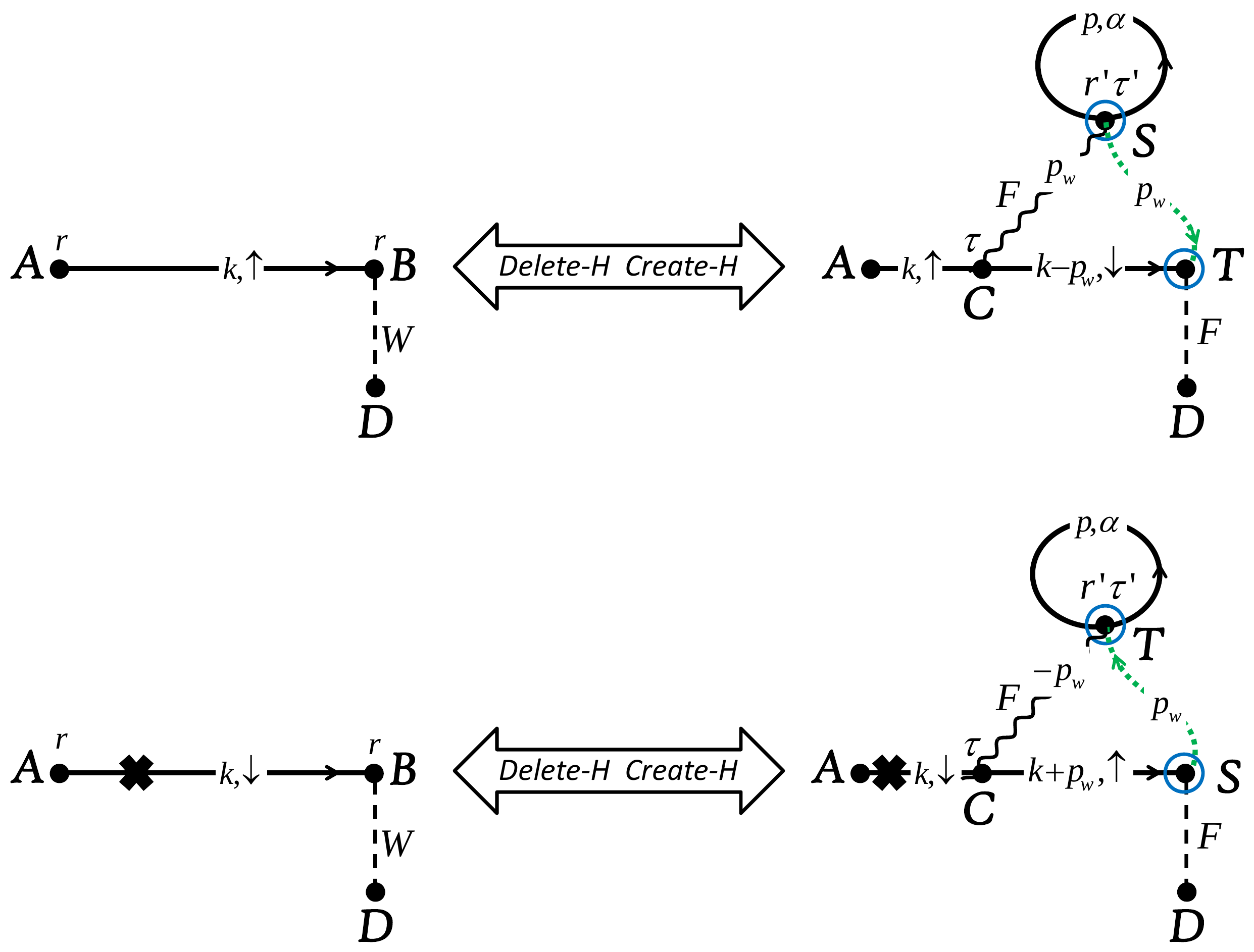}
\caption{\label{fig:6}
  Two cases for {\it Create-H} and {\it Delete-H} updates which
  insert/remove a pair of worms and increase the diagram order by
  adding a Hartree-type bubble.
}
\end{figure}

\paragraph*{\underline{Create-H - Delete-H}}
This pair of complementary updates also switches between physical and
unphysical sectors with an additional ingredient---it
increases/decreases the diagram order by attaching a Hartree-type
bubble to the existing graph. According to the rule 'no two lines may
have the same momentum' the diagram remains irreducible because one of
the worms is placed on the bubble vertex. The overall transformation
is illustrated in Fig.~\ref{fig:6}; the text below addresses to this
figure with regards to the procedure of selecting specific graph
parameters.  In {\it Create-H} a particle propagator (going from
vertex ${\cal A}$ to vertex ${\cal B}$) and whether to place ${\cal
  S}$ or ${\cal T}$ on vertex ${\cal B}$ is decided at random. If the
proposal is inconsistent with the worm rules it has to be
rejected. Next, a new time variable for the intermediate vertex ${\cal
  C}$ is generated from the probability density, $t(\tau )$, and a
random decision is made whether the new interaction line is of the
$J$- or $\tilde{W}$-type. For the $J$-line (with $\tau ' = \tau $) the
position of the second worm vertex in space is obtained from the
normalized $X(r') \propto J(r')$ distribution.  For the
$\tilde{W}$-line this position is obtained from some designed
probability distribution $Y(r')$ while the time location is drawn from
the probability density, $t (\tau ')$. The spin variable in the
bubble, the bubble momentum variable $p$ and the worm momentum $p_w$
are decided at random.  In {\it Delete-H} a random choice is made what
type of special vertex must be on the Hartree bubble provided the
overall topology of lines is identical to that on the r.h.s of
Fig.~\ref{fig:6}.  The proposal is to remove worms and the bubble from
the diagram. It is rejected if either the propagator originating from
${\cal C}$ or the interaction line attached to ${\cal C}$ is the dummy
one.  The acceptance ratios for these updates are (probabilities of
calling {\it Create-H} are {\it Delete-H} are $u_3$ and $u_4$,
respectively)
\begin{equation}
R_{\rm Create-H}= \frac{D_{\nu '}}{D_{\nu}}  \frac{u_4}{u_3} \frac{2^3n}{t(\tau)}
\left\{
\begin{array}{ll}
 1/X(r')           ; & \;\; (J)  \\
 1/Y(r') t(\tau ') ; & \;\; (\tilde{W})
\end{array}
\right.
\label{U34a}
\end{equation}
\begin{equation}
R_{\rm Delete-H}= \frac{D_{\nu '}}{D_{\nu}}  \frac{u_3}{u_4} \frac{t(\tau)}{2^3(n-1)}
\left\{
\begin{array}{ll}
 X(r') ;           & \;\; (J)  \\
 Y(r') t(\tau ') ; & \;\; (\tilde{W})
\end{array}
\right.
\label{U34b}
\end{equation}
with the diagram weight ratios $D_{\nu '}/D_{\nu}$ given by
$|GGFFG/GW|$ and $|GW/GGFFG|$ in {\it Create-H} and {\it Delete-H},
respectively (note that here $n$ is the initial diagram order).
The simplest choices for probability distributions in Eqs.(\ref{U34a}) and (\ref{U34b}) would be
uniform distributions $t(\tau )=T$, $X(r')=1/z$, and $Y(r')=1/V$, where $V$
is the total number of lattice sites. One can use other functions
for better acceptance ratio.
%
The three updates described next represent a random diffusion of
special vertexes along the graph lines (this places ${\cal S}$ and
${\cal T}$ on any allowed pair of vertexes) supplemented by an update
which changes the graph topology.

\begin{figure}[htbp]
\includegraphics[angle=0,width=1.\columnwidth]{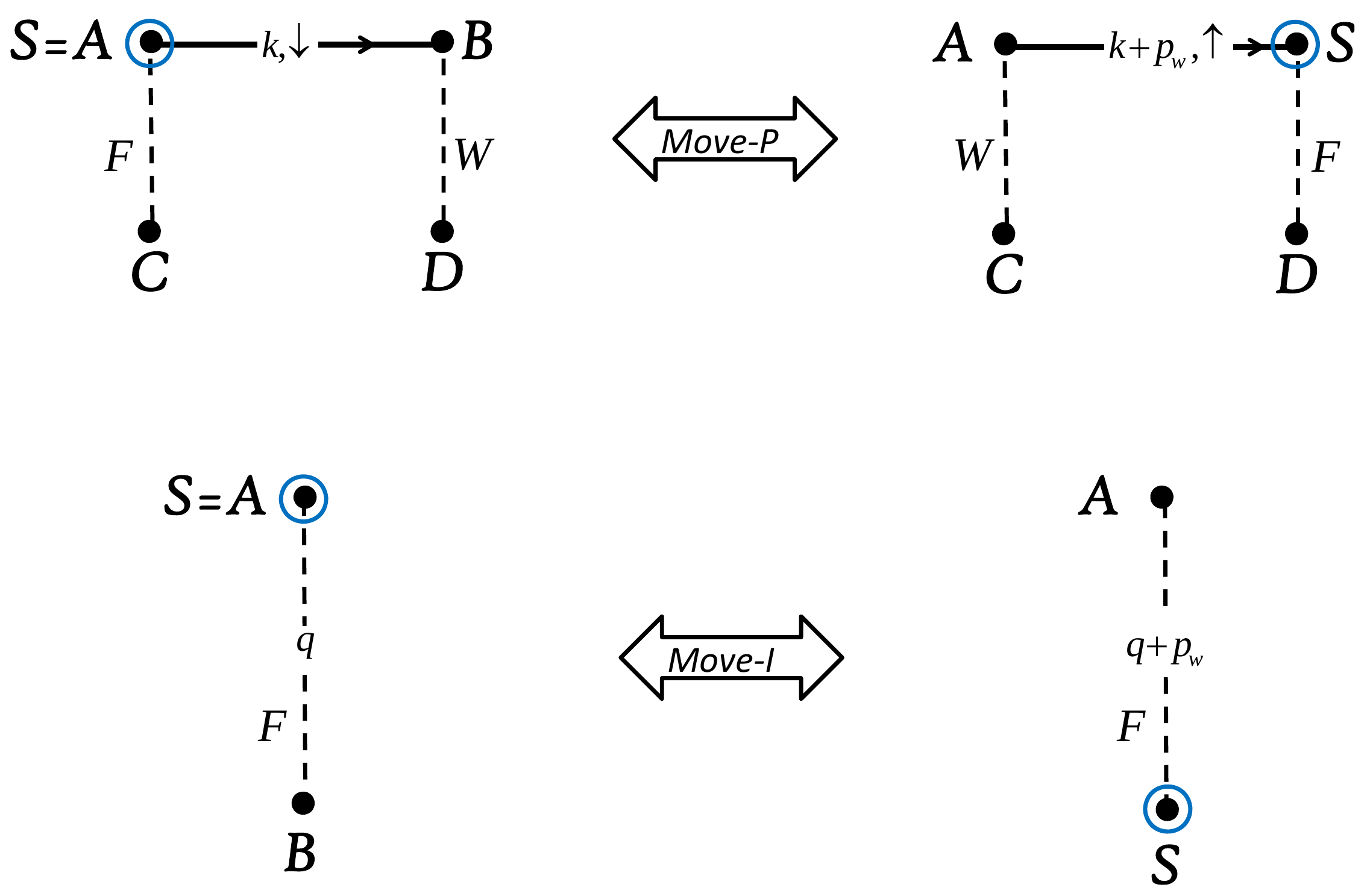}
\caption{\label{fig:7}
  Upper panel: An illustration of the {\it Move-P} update shifting
  ${\cal S}$ along the propagator line. This move changes the momentum
  of the propagator line and flips its spin as well as the status of interaction lines
  between physical and unphysical.  Lower panel: An illustration of
  the {\it Move-I} update shifting ${\cal S}$ along the interaction
  line. This move changes the auxiliary momentum of the interaction
  line only.
}
\end{figure}

\paragraph*{\underline{Move-P}}
In this self-complementary update one selects at random ${\cal S}$ or
${\cal T}$ and proposes to shift the selected worm along the incoming
or outgoing propagator line, deciding again randomly between the two
choices. Since all four case are identical in their implementation we
describe below an update shifting ${\cal S}$ along the outgoing
propagator line to vertex {\cal B}, see upper panel in
Fig.~\ref{fig:7}, if ${\cal B} \ne {\cal S}$, of course.  According to
the rules, if ${\cal B}={\cal T}$, or ${\cal D}={\cal T}$, or the spin
of the outgoing line is down, the update is rejected. Finally, one has
to check whether the proposal may lead to the irreducible diagram.
For the update shown in Fig.\ref{fig:7} the acceptance ratio is given by
\begin{equation}
R_{\rm Move-P}= \left\vert \frac{GWF}{GFW} \right\vert \;.
\label{U5}
\end{equation}

\paragraph*{\underline{Move-I}}
In this self-complementary update one selects at random ${\cal S}$ or
${\cal T}$ and proposes to shift the selected worm along the
interaction line to vertex ${\cal B}$, see the lower panel in
Fig.~\ref{fig:7}.  One has to check whether the proposal may lead to
the irreducible diagram.  This update is always accepted since it
changes only the auxiliary momentum variable.

\begin{figure}[htbp]
\includegraphics[angle=0,width=1.\columnwidth]{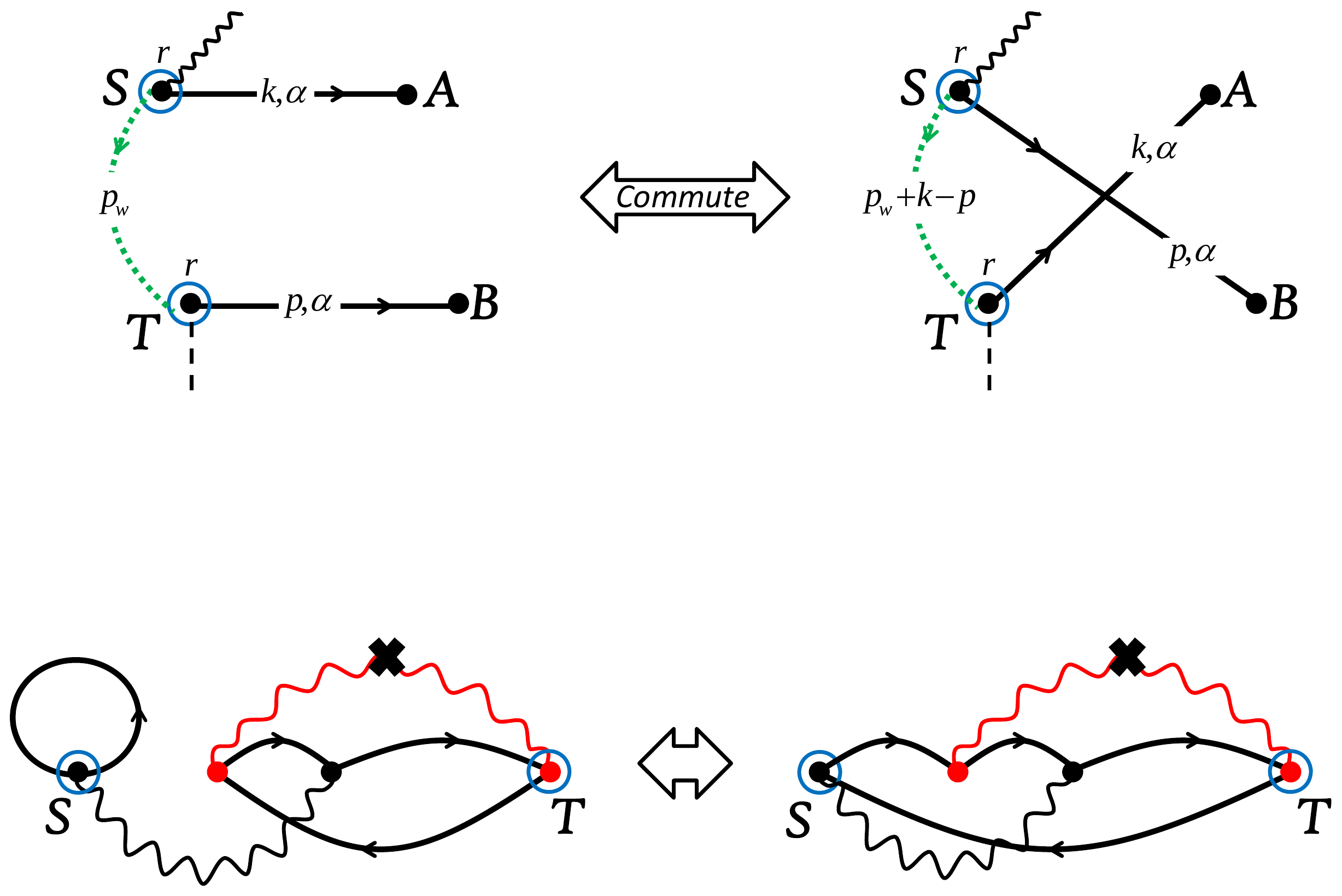}
\caption{\label{fig:8}
  Upper panel: In the {\it Commute} update the diagram topology is
  changed by re-directing propagators originating at special vertexes
  ${\cal S}$ and ${\cal T}$ to have their destination vertexes at
  ${\cal B}$ and ${\cal A}$, respectively. Lower panel: typical
  transformation produced by the {\it Commute} update.
}
\end{figure}

\paragraph*{\underline{Commute}}
All possible topologies in a graph of order $n$ can be generated by
randomly connecting outgoing propagator lines to incoming ones. Having
this observation in mind the self-complementary {\it Commute} update
proposes to swap destination vertexes for propagator lines originating
at ${\cal S}$ or ${\cal T}$, see Fig.~\ref{fig:8}. This proposal is
valid only if all vertexes have the same space coordinate and both
propagators have the same spin index. The crucial advantage of the
worm algorithm at this point becomes clear---momentum conservation
law is satisfied by absorbing the difference $k-p$ into the worm
momentum $p_w \to p_w+k-p$. This update is accepted with ratio
\begin{equation}
R_{\rm Commute}= \left\vert \frac{GG}{GG} \right\vert  \;.
\label{U7}
\end{equation}
In the lower panel of Fig.~\ref{fig:8} we show a typical diagram
change produced by {\it Commute}. It always changes the number of
fermionic loops in the graph and, in particular, will transform a
diagram with a bubble attached to the worm vertex into the
vertex-correction type diagram. This explains, to some extent, our
design of updates {\it Create-H} and {\it Delete-H}.
%

\paragraph*{\underline{Dummy}}
To place the dummy line mark on any of the $G$- or $\tilde{W}$-lines
we select one of the vertexes at random, say vertex ${\cal A}$, and
make a random decision whether the new dummy line should be the
interaction line attached to ${\cal A}$ (it has to be of the
$\tilde{W}$ type) or the propagator line originating from ${\cal
  A}$. The proposal is always accepted if the notion of the dummy mark
does not change the value of the function behind the line (which is
assumed to be the case here).

The above set of updates is sufficient for doing the simulation.  It
can always be supplemented by additional updates which do not
necessarily involve special vertexes but reduce the autocorrelation
time and lead to more efficient sampling of the diagram variables.
Moreover, a non-trivial check of the detailed balance for debugging
purposes is only possible if the set of updates is overcomplete. Below
we describe several such updates.

\begin{figure}[htbp]
\includegraphics[angle=0,width=1.\columnwidth]{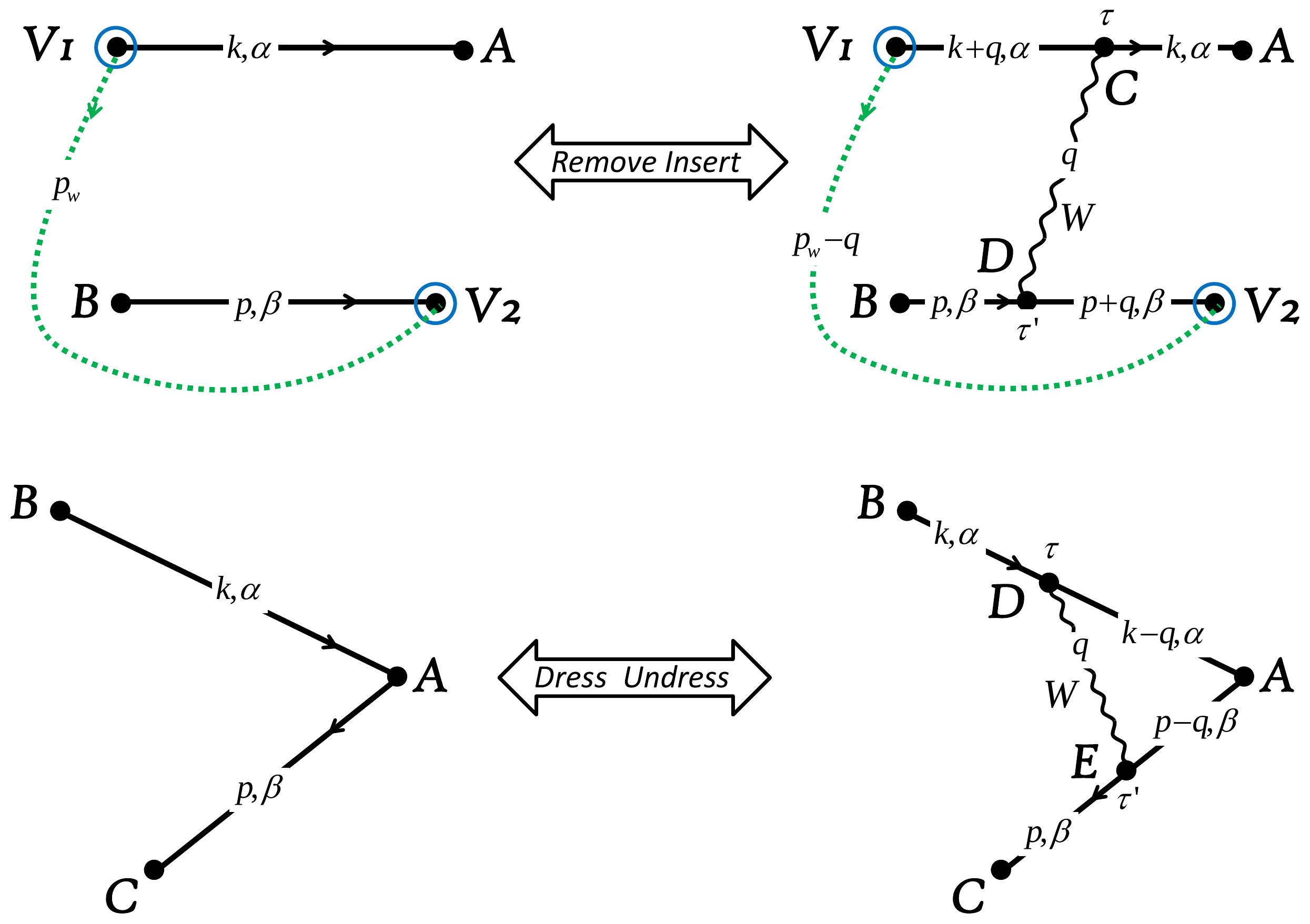}
\caption{\label{fig:9}
  Upper panel: Increasing/decreasing the diagram order using a pair of
  complementary updates {\it Insert} and {\it Remove}. When the
  propagators ${\cal V}_1$-${\cal A}$ and ${\cal B}$-${\cal V}_2$ are
  linked with the new interaction line ${\cal C}$-${\cal D}$ carrying
  momentum $q$ the closed loop for momentum conservation goes as
  ${\cal V}_1$-${\cal C}$-${\cal D}$-${\cal V}_2$-${\cal V}_1$. The
  same loop is used in the {\it Remove} update. Lower panel: Diagram
  transformation when vertexes are dressed and undressed with
  interaction lines.
}
\end{figure}

\paragraph*{\underline{Insert-Remove}}
An idea here is to increase/decrease the diagram order by
inserting/removing a ladder-type structure. More precisely, in {\it
  Insert} we make a random choice between ${\cal S}$ and ${\cal T}$ to
start the construction from the special vertex ${\cal V}_1$. Next we
identify vertex ${\cal A}$ as the destination vertex of the propagator
originating from ${\cal V}_1$, and vertex ${\cal B}$ as the
originating vertex for the propagator with the destination vertex
${\cal V}_2$, which is the other worm end, see the upper panel in
Fig.~\ref{fig:9}. If ${\cal A}={\cal V}_2$ the update is rejected. The
proposal is to insert new vertexes ${\cal C}$ (intermediate between
${\cal V}_1$ and ${\cal A}$) and ${\cal D}$ (intermediate between
${\cal B}$ and ${\cal V}_2$) and to link them with the interaction
line of randomly chosen type, either $J$ or $\tilde{W}$, and momentum
$q$.  The new time variables are generated from the probability
density $t(\tau )$. The momenta of new lines and the worm are modified
as described in Fig.~\ref{fig:9} to satisfy conservation laws.
Finally, if one of the propagators is the dummy line, the new dummy
line has to have the same originating vertex. In {\it Remove} we
select one of the special vertexes, ${\cal V}_1$, at random and verify
that the topology of lines connecting it to the other special vertex,
${\cal V}_2$, as well as lines parameters are consistent with
Fig.~\ref{fig:9} (upper panel). If either ${\cal C}$-${\cal A}$ or
${\cal D}$-${\cal V}_2$ propagator is a dummy line the update is
rejected. The proposal then is to remove vertexes ${\cal C}$ and
${\cal D}$ from the graph and update momenta of the lines
accordingly. The acceptance ratios are given by
\begin{equation}
R_{\rm Insert}= \frac{D_{\nu '}}{D_{\nu}}  \frac{u_{10}}{u_9} \frac{2}{t(\tau)}
\left\{
\begin{array}{ll}
 1           ; & \;\; (J)  \\
 1/t(\tau ') ; & \;\; (\tilde{W})
\end{array}
\right.
\label{U910}
\end{equation}
\begin{equation}
R_{\rm Remove}= \frac{D_{\nu '}}{D_{\nu}}  \frac{u_{9}}{u_{10}} \frac{t(\tau)}{2}
\left\{
\begin{array}{ll}
 1           ; & \;\; (J)  \\
 t(\tau ') ; & \;\; (\tilde{W})
 \end{array}
\right.
\label{U910b}
\end{equation}
where $D_{\nu '}/D_{\nu} = |GGGGW/GG|$ in {\it Insert} and its inverse in {\it Remove}.
%

\paragraph*{\underline{Dress-Undress}}
One of the easiest updates to increase the diagram order within the
$G^2W$-skeleton formulation is to dress an existing vertex with
interaction line and consider the smallest closed loop for
transferring momentum. The {\it Dress} update starts from random
selection of vertex ${\cal A}$ and identification of vertexes ${\cal
  B}$ and ${\cal C}$ linked to it by propagator lines; if ${\cal
  B}={\cal A}$ the update is rejected (we pay no attention in this
update whether one of the vertexes is of a special type). The proposal
is to add new vertexes ${\cal D}$ (intermediate between ${\cal B}$ and
${\cal A}$) and ${\cal E}$ (intermediate between ${\cal A}$ and ${\cal
  C}$) and to link them with the $\tilde{W}$ line with random momentum
$q$. The new time variables are generated from the probability density
$t(\tau )$. The momenta of new lines are modified as described in the
lower panel of Fig.~\ref{fig:9} to satisfy conservation laws. In {\it
  Undress} we select vertex ${\cal A}$ at random, identify vertexes
${\cal D}$, ${\cal B}$, ${\cal E}$, and ${\cal C}$ using links along
the propagator lines, and verify that the topology of lines and their
parameters are consistent with the dressed vertex configuration. If
the ${\cal D}$-${\cal E}$ line is not of the diagonal $\tilde{W}$ type
or one of the propagators ${\cal D}$-${\cal A}$ or ${\cal E}$-${\cal
  C}$ is a dummy line, the update is rejected. The proposal is to
remove vertexes ${\cal D}$ and ${\cal E}$ from the graph. The
acceptance ratios are
\begin{equation}
R_{\rm Dress}= \frac{D_{\nu '}}{D_{\nu}}  \frac{u_{12}}{u_{11}} \frac{n}{(n+1)t(\tau)t(\tau ')} \;,
\label{U1112}
\end{equation}
\begin{equation}
R_{\rm Undress}= \frac{D_{\nu '}}{D_{\nu}}  \frac{u_{12}}{u_{11}} \frac{n}{(n-1)t(\tau)t(\tau ')} \;,
\label{U1112b}
\end{equation}
where $D_{\nu '}/D_{\nu} = |GGGGW/GG|$ in {\it Dress} and its inverse in {\it Undress}.

\paragraph*{\underline{Recolor}}
An easy and efficient way to change the spin index of propagator lines
is to select a random vertex and use it to construct a closed loop by
following the propagator lines attached to it. If all propagators in
the loop have the same spin index $\alpha $ it can be changed to
$-\alpha$ with acceptance ratio unity in the absence of external
magnetic field; otherwise, one has to use the ratio of products of
all propagator lines after and before the update.
%
\begin{figure}[htbp]
\includegraphics[angle=0,width=1.\columnwidth]{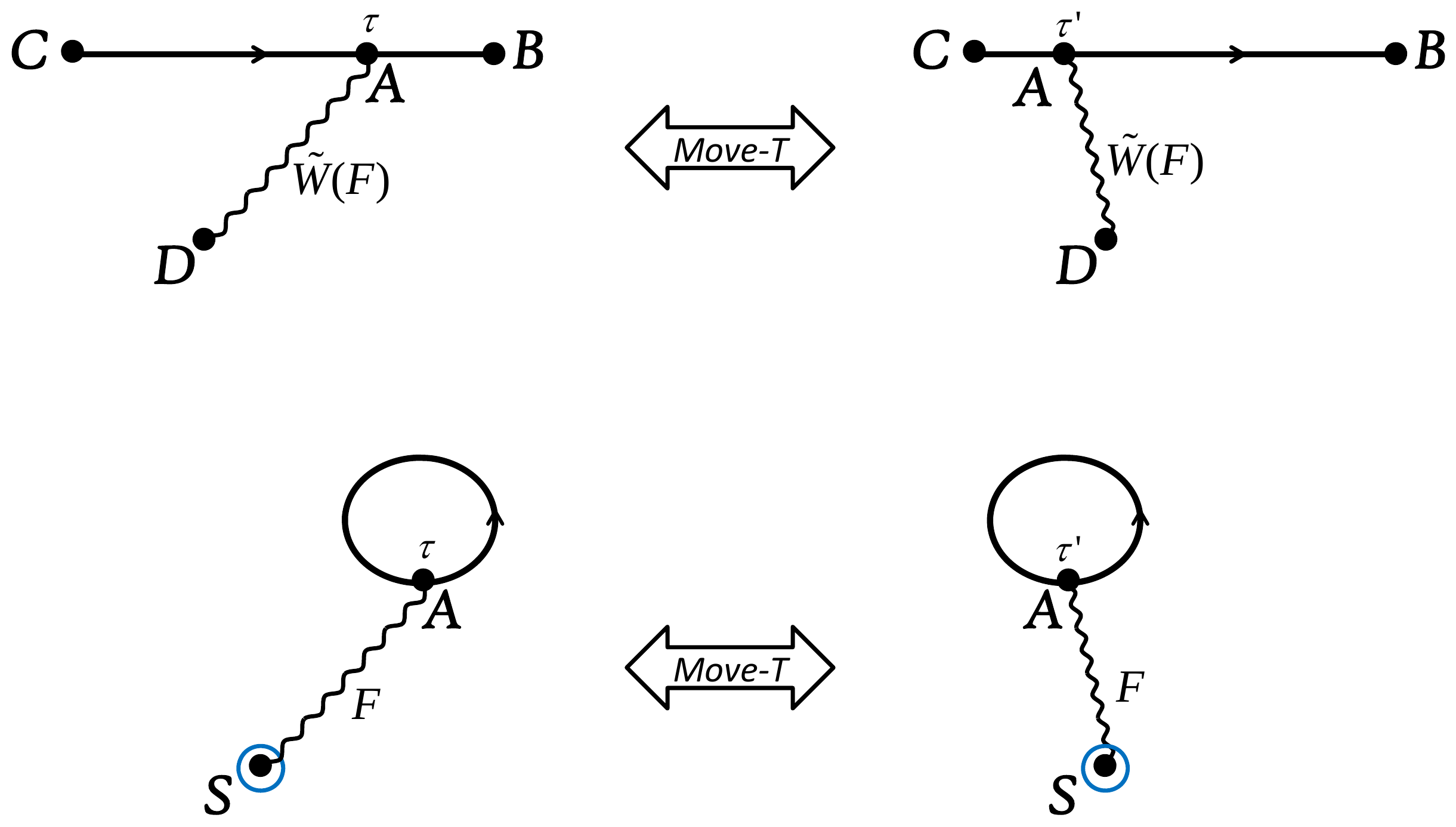}
\caption{\label{fig:10}
  An illustration of the {\it Move-T} update changing the imaginary
  time location of a randomly chosen vertex. One is free to change the
  type of worm from ${\cal S}$ to ${\cal T}$ or to place it on any of
  the vertexes in both panels.
}
\end{figure}

\paragraph*{\underline{Move-T}}
This self-complementary update is designed to sample time variables of
the diagram without changing its order and topology. The proposal is
to select one of the vertexes at random, let it be vertex ${\cal
  A}$ and update its imaginary time variable from $\tau $ to $\tau '$
using probability density distribution $t(\tau ')$, see
Fig.~\ref{fig:10} for two alternatives.  The interaction line attached
to ${\cal A}$ cannot be of the $J$-type; whether it is the physical
$\tilde{W}$-line or unphysical $F$-line does not mater. The acceptance
ratio is given by
\begin{equation}
  R_{\rm Move-T}= \frac{t(\tau )}{t(\tau ')}
  \begin{cases}
    |\tilde{W}(F)/\tilde{W}(F)|\, |GG/GG| & \text{(generic)}\\
    |F/F| & \text{(bubble)}
  \end{cases}.
\label{U14}
\end{equation}

\subsection{Diagram sign}
With the dummy line removed, the diagram phase required to compute the
self-energy and polarization operator using Eqs.~\eqref{D2} and
\eqref{N1} is determined by standard diagrammatic rules (see also
Eq.~\eqref{D0}):
\begin{equation}
\varphi_{\nu} = \sum_{\rm lines} \arg (f_{\rm line}(\nu )) + \pi (n+l) \;,
\label{phase}
\end{equation}
where $l$ is the number of fermionic loops (one only needs to know
whether it is even or odd).  This phase is readily recalculated in
updates without addressing the whole diagram since $l$ always changes
its parity when {\it Create-H}, {\it Delete-H}, and {\it Commute} are
accepted.

\subsection{Satisfying the sum rule}

The value of the spin-spin correlation function $\chi (r=0, \tau =0)
=\langle (S^z)^2 \rangle= 1/4$, see Eq.~\eqref{chi}, provides an important
sum rule in the Fourier space
\begin{equation}
  T\sum_n \int_{BZ} \frac{d{\mathbf q}}{8\pi^2/\sqrt{3}} \chi
  (q, n ) = 1/4 \;,
\label{sumrule}
\end{equation}
which can be used for modifying convergence properties of the
self-consistent scheme as follows
(the integral is taken over the Brillouin zone (BZ)). When the maximum diagram order is
fixed at $N$ the sum rule is violated by some amount which vanishes as
$N\to \infty$. Since the final result is claimed after taking the
limit, it is perfectly reasonable to impose a condition that the sum
rule is always satisfied by scaling $\Pi$ by an appropriate factor.
This is exactly what is done in this article: after solving the Dyson
Equation we check the value of $\chi (r=0, \tau =0)$, adjust the
scaling factor for $\Pi$, and go back to solving the Dyson Equation
again until the sum rule is satisfied with three digit accuracy.

\section{Triangular lattice Heisenberg antiferromagnet}
\label{sec:5}

In the diagrammatic formulation there is no conceptual difference in
the implementation of the numerical scheme for any dimension of space,
lattice type, and interaction range. Thus, sign-problem free systems,
e.g. the square/cubic lattice Heisenberg antiferromagnet with nearest
neighbor coupling, can be used for testing purposes since their
properties are known with high degree of accuracy (with reliable
extrapolation to the thermodynamic limit) using path-integral and
stochastic series expansion MC methods. After passing such tests, we
turn our attention to the triangular-lattice Heisenberg
antiferromagnet (TLHA) which is a canonical frustrated magnetic system with
massively degenerate ground state in the Ising limit.

\begin{figure}[htbp]
\includegraphics[angle=0,width=0.9\columnwidth]{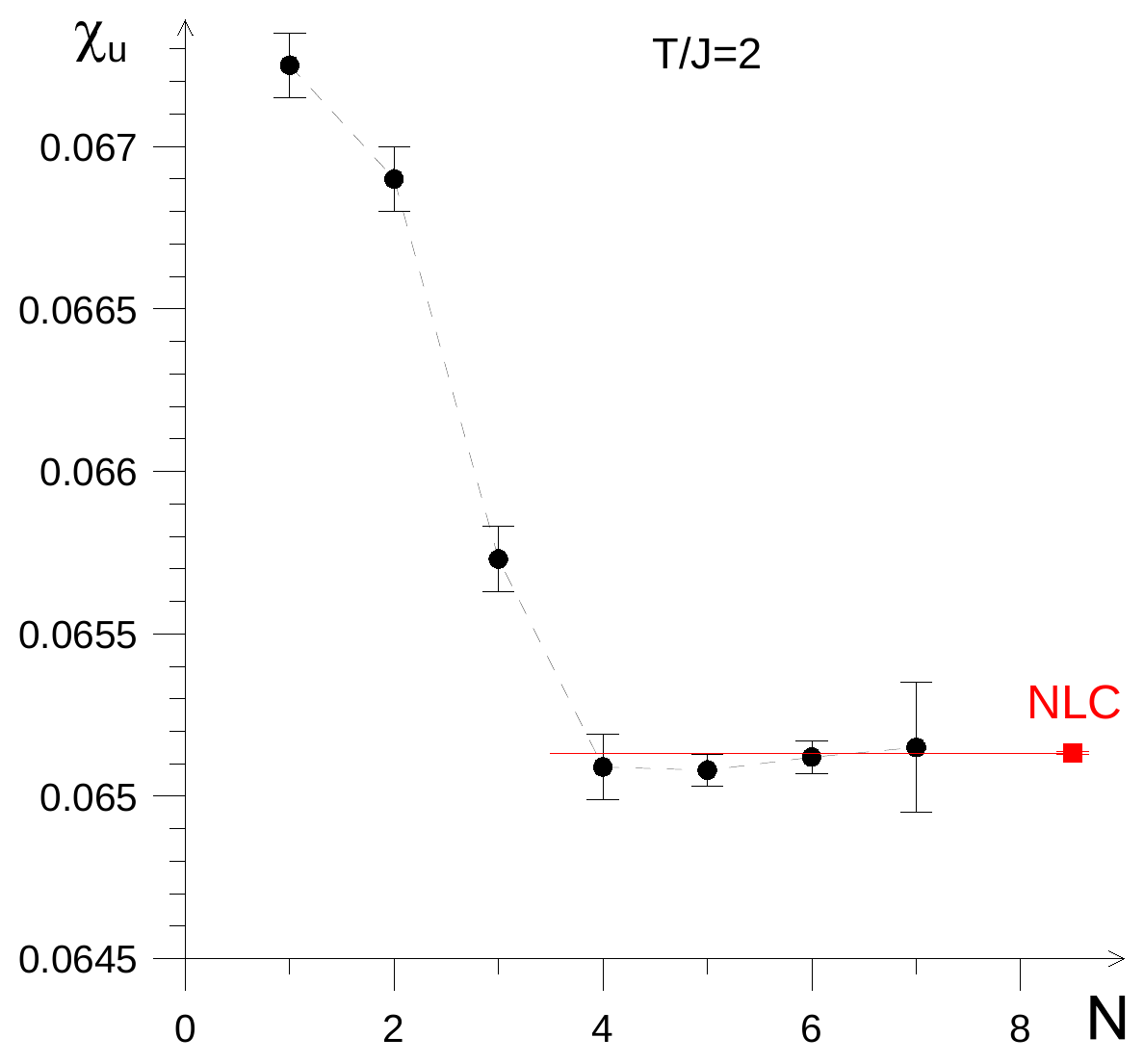}
\caption{\label{fig:11}
  (Color online) Uniform susceptibility calculated within the
  $G^2W$-skeleton expansion as a function of the maximum diagram order
  retained in the BDMC simulation (black dots) for $T/J=2$. The result
  of the high-temperature expansion (with Pad\'e approximant extrapolation)
  \cite{singh05} is shown by red square and horizontal line.
}
\end{figure}

The most important question to answer is whether the sign blessing
phenomenon indeed takes place, i.e. there is a hope for obtaining
accurate predictions in the strong coupling regime by calculating
higher-and-higher order diagrams despite factorial growth in the
number of contributing graphs. In Fig.~\ref{fig:11} we show comparison
between the calculated answer for the static uniform susceptibility
\begin{equation}
\chi_u =\chi(q=0,m=0)= \int_0^{1/T} d\tau \sum_{r}\chi (r,\tau )   \, ,
\label{chiu}
\end{equation}
and the high-temperature expansion results~\cite{singh05,Rigol} at $T/J=2$. This temperature
is low enough to ensure that we are in the regime of strong
correlations because $\chi_u$ is nearly a factor of two smaller than
the free spin answer $\chi_u^{0} = 1/4T$.  On the other hand, this
temperature is high enough to be sure that the high-temperature series can be
described by Pad\'e approximants without significant systematic
deviations from the exact answer~\cite{singh05,Rigol} (at slightly lower
temperature the bare NLC series start to diverge).  We clearly see in
Fig.~\ref{fig:11} that the BDMC series converges to the correct result
with accuracy of about three meaningful digits and there is no
statistically significant change when more than a hundred thousand of
7-th order diagrams are accounted for.  [We recall that the number of topologically
distinct diagrams within the $G^2W$-skeleton scheme was calculated in
Ref.~\onlinecite{molinari}; for the eight lowest orders they are
$1,1,6,49, 542, 7278, 113824, 2017881$.]  The error bar for the 7-th
order point is significantly increased due to exponential growth in
computational complexity. The 4-th order result can be obtained after
several hours of CPU time on a single processor.
%

\begin{figure}[htbp]
\includegraphics[angle=0,width=0.9\columnwidth]{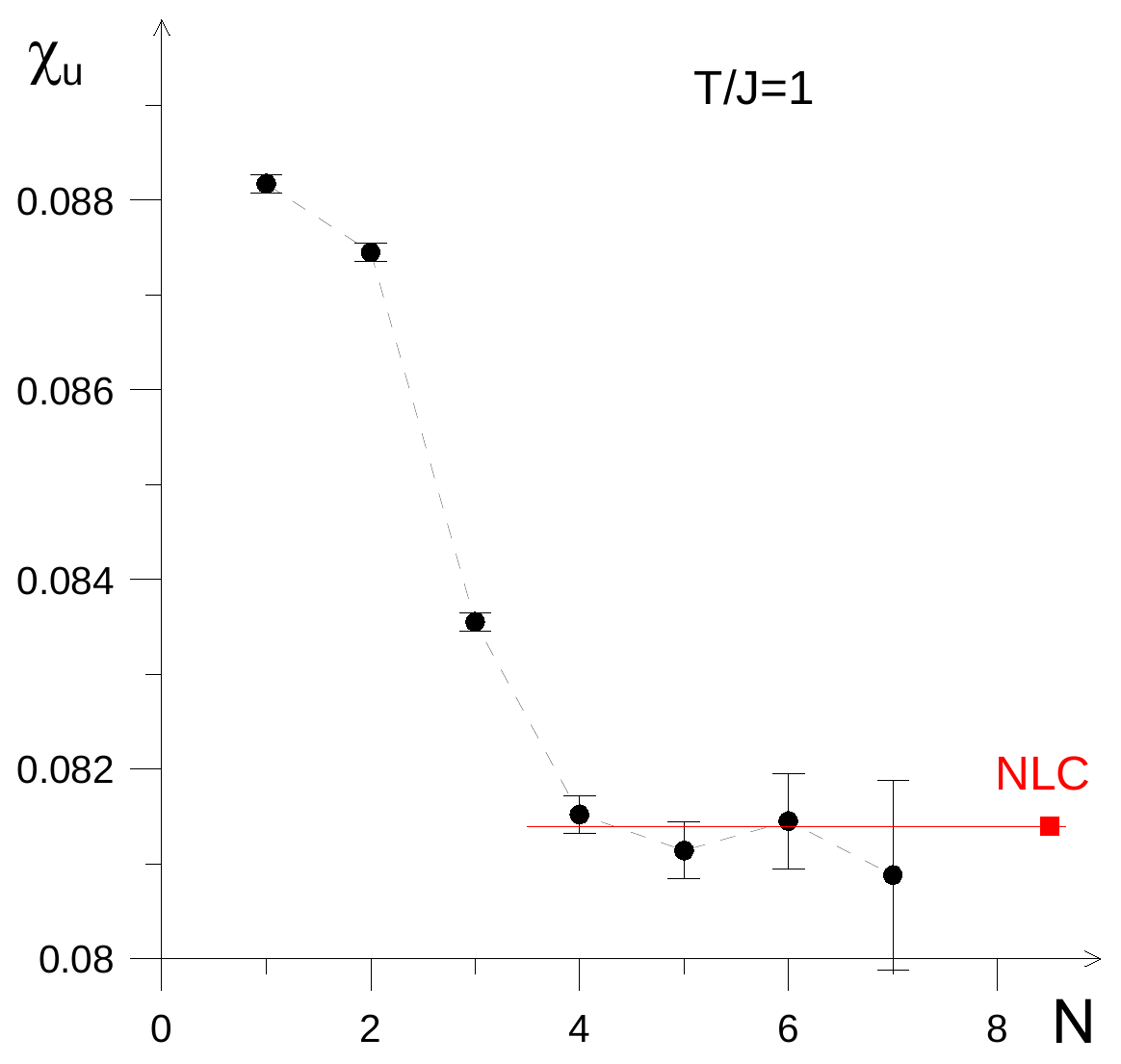}
\caption{\label{fig:12}
  (Color online) Uniform susceptibility as a function of the maximum
  diagram order (black dots) for $T/J=1$. The result of the high-temperature
  expansion (with Pad\'e approximant extrapolation) \cite{singh05} is shown by red
  square and horizontal line. Its error bar is based on the difference
  between various expansion/extrapolation schemes.
}
\end{figure}

\begin{figure}[htbp]
\includegraphics[angle=0,width=0.9\columnwidth]{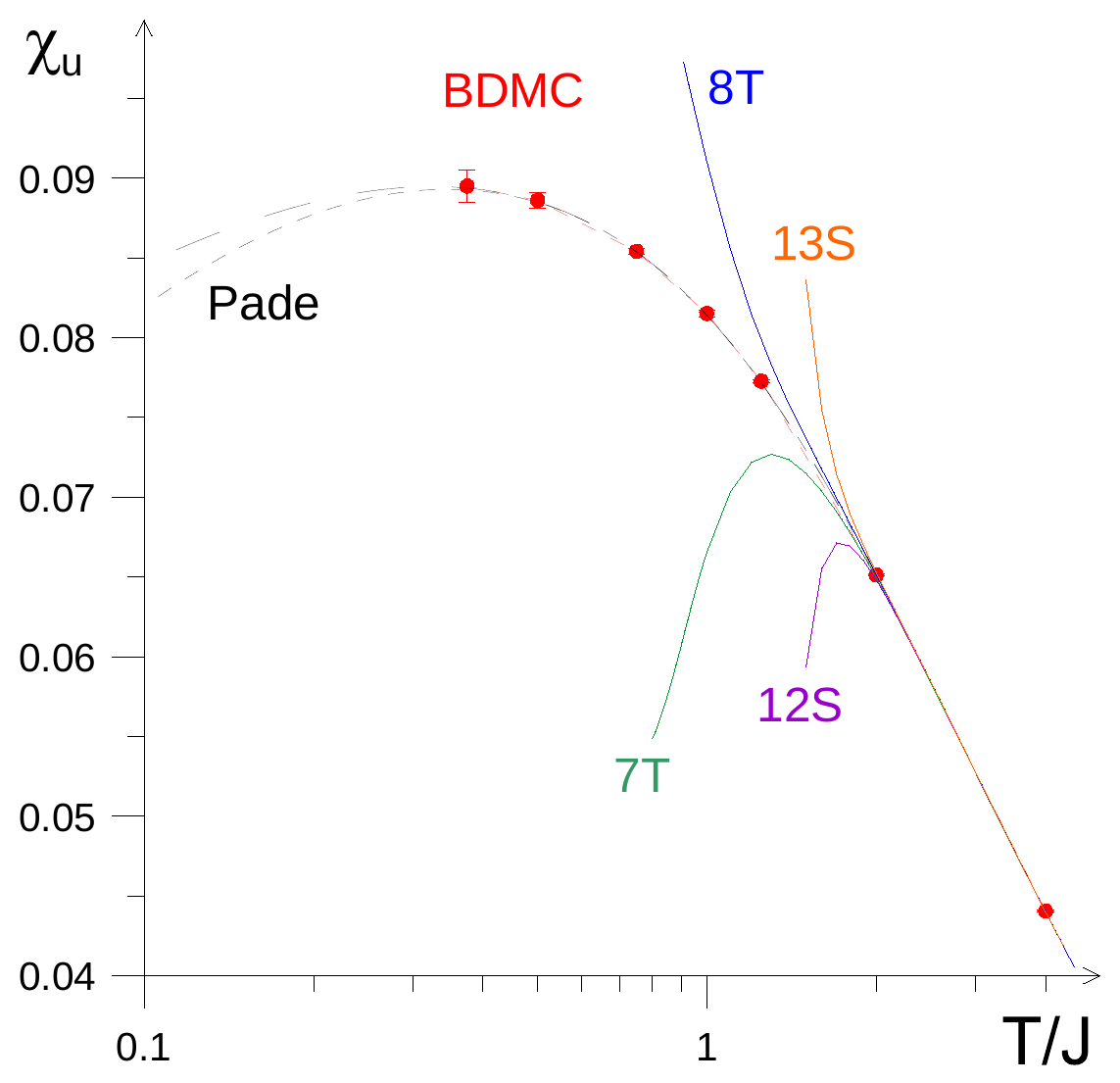}
\caption{\label{fig:13}
  (Color online) Uniform susceptibility as a function of temperature
  (red dots) for the triangular Heisenberg antiferromagnet
  calculated within the BDMC approach.  NLC expansion results ~\cite{Rigol} based on
  triangles (labeled as 7T and 8T) and sites (labeled as 12S and 13S)
  are shown along with two different Pad\'e approximant
  extrapolations of high-temperature expansions~\cite{singh05}.
}
\end{figure}

Interestingly enough, when temperature is lowered down to $T/J=1$,
which is significantly below the point where the bare NLC series start
to diverge (see Fig.~\ref{fig:13}), the BDMC series continue to
converge (see Fig.~\ref{fig:12}). This underlines the importance of
performing simulations within the self-consistent skeleton
formulation.
%

In Fig.~\ref{fig:13} we show results of the BDMC simulation performed
at temperatures significantly below the mean-field transition
temperature. For all points we observe extremely
good agreement (essentially within our error bars) with the Pad\'e
approximants used to extrapolate the high-temperature expansion data
to lower temperature \cite{singh05}.
Within the current protocol of dealing with skeleton diagrams we were
not able to go to lower temperature due to the development of
singularity in the response function (and thus effective interaction
$\tilde{W}$ at the wave-vector $Q=(4\pi/(3a),0)$. When the denominator
$4 - J(q) \Pi (q,m)$ in \eqref{dysonGW} is close to zero it becomes very difficult to control
highly non-linear sets of coupled integral equations given finite
statistical noise on the measured quantity $\Pi (q,m)$.

\begin{figure}[htbp]
\includegraphics[angle=0,width=0.9\columnwidth]{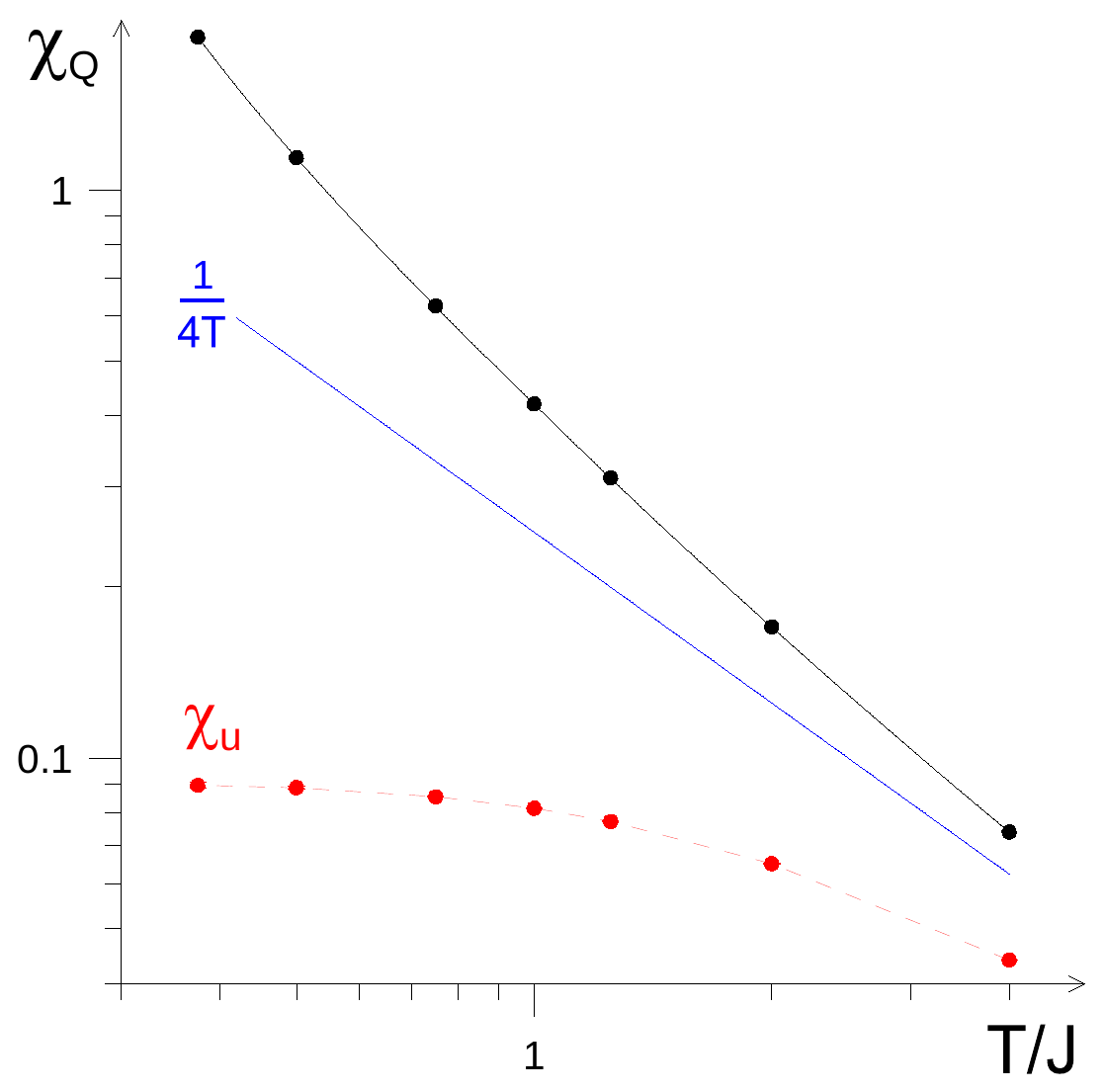}
\caption{\label{fig:14}
  (Color online) Staggered susceptibility at the wave vector $Q$
  as a function of temperature (black dots) plotted for
  comparison along with the Curie-Weiss law (blue curve) and uniform
  susceptibility (red dots and line).
}
\end{figure}

This is clearly seen in Fig.~\ref{fig:14} where we show data for the
staggered susceptibility $\chi(Q,0)$, defined as
\begin{equation}
\chi_s =\chi(Q,m=0)= \int_0^{1/T} d\tau \sum_r \; e^{iQ\cdot r } \chi (r,\tau )  \, ,
\label{chis}
\end{equation}
along with the Curie law and the uniform susceptibility, on the double
logarithmic scale.
%
%

\begin{figure}[tbp]
\includegraphics[angle=0,width=0.9\columnwidth]{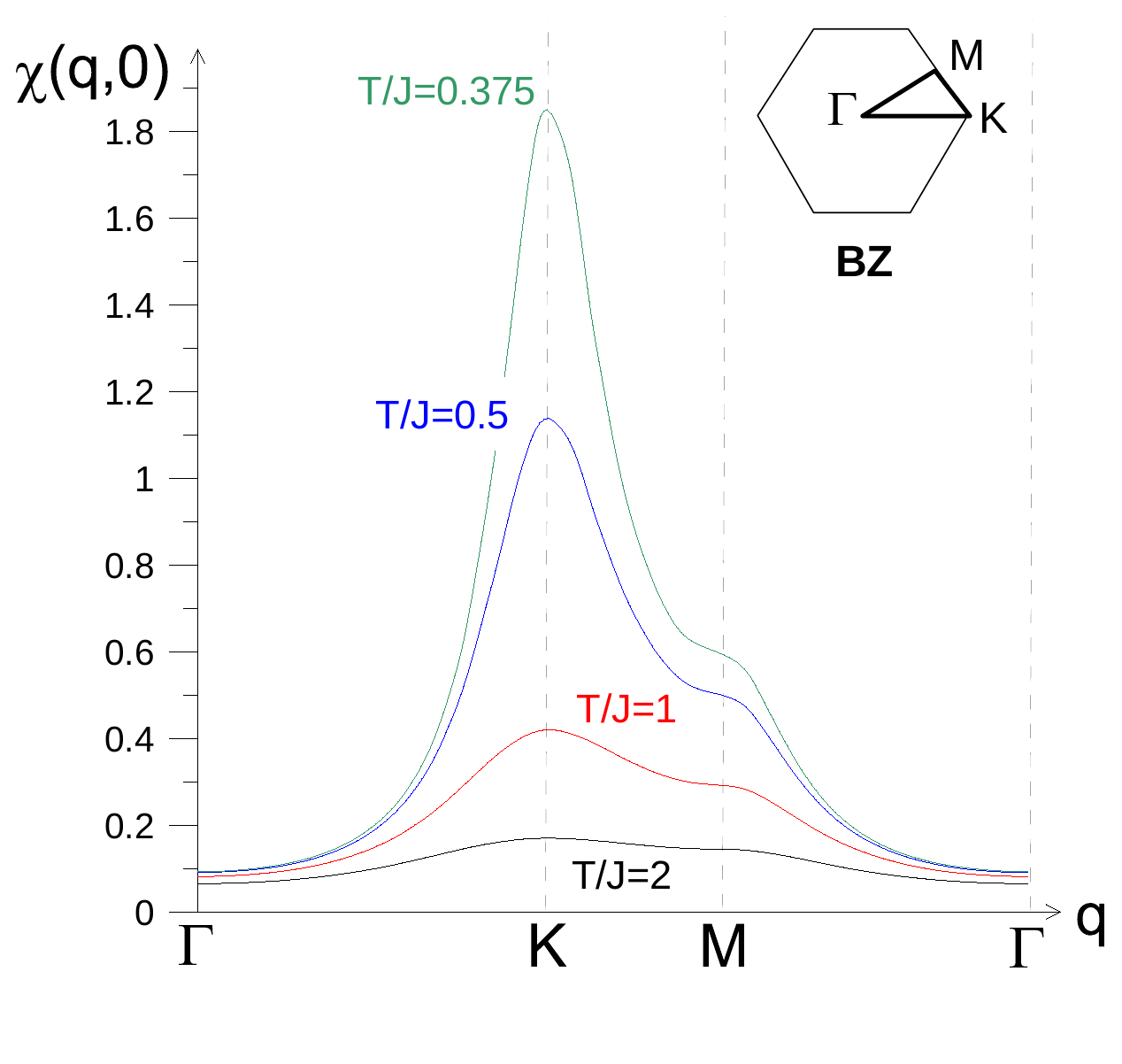}
\caption{\label{fig:15}
  (Color online) Static spin-spin correlation function along the
  characteristic trajectory in the Brillouin zone.
}
\end{figure}

One of the advantages of our approach is the ability to perform
calculations of susceptibility at arbitrary momentum.  In
Fig.~\ref{fig:15} we show data for $\chi(q, 0)$ along the
$\Gamma - K - M - \Gamma $ trajectory in the Brillouin zone (BZ). Here
$\Gamma$ is the center of the BZ, $K = Q=(4\pi/(3a),0)$, and
$M = (\pi/a, \pi/(\sqrt{3} a))$ is the mid-point on the face of the
hexagonal BZ (see Fig.~\ref{fig:15}).  Results presented in
Figs.~\ref{fig:14} and \ref{fig:15} are new because they are obtained
for the static (zero Matsubara frequency) susceptibility in the
thermodynamic limit. It is useful to note that the static response is
far more difficult to get within the NLC method which is suited for
calculations of the equal time correlation functions, such as, for
example, equal time spin structure factor. The exception is
represented by the uniform, or zero momentum, response which is based
on the total magnetization commuting with the Hamiltonian. In
addition, we can afford very high resolution in momentum space which
is not the case for calculations based on clusters of finite size.
%

\begin{figure} 
%
%
\begin{tabular}{lr}
\includegraphics[angle=0,width=0.65\columnwidth]{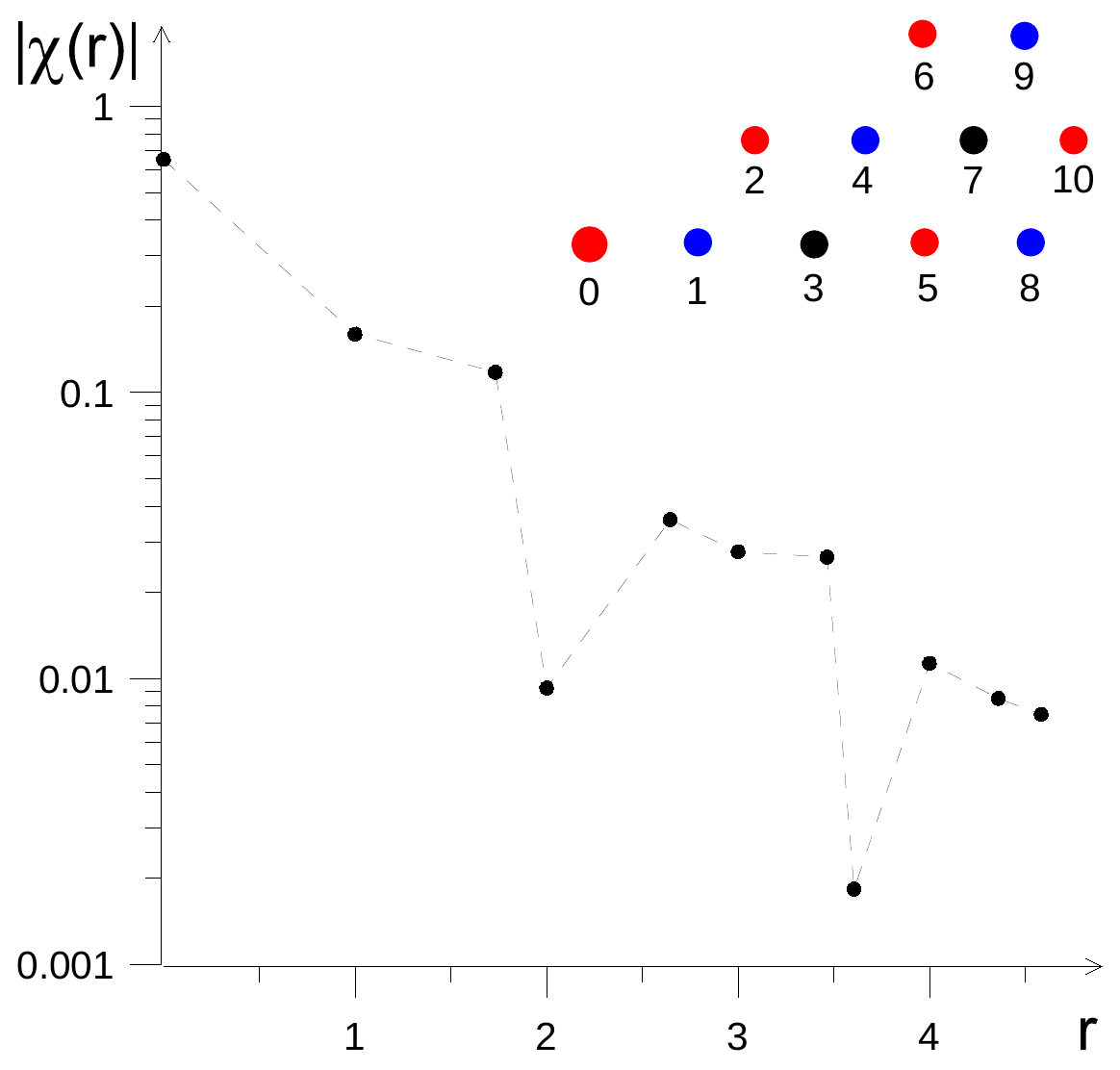}&
\raisebox{0.34\columnwidth}{\includegraphics[angle=0,width=0.34\columnwidth]{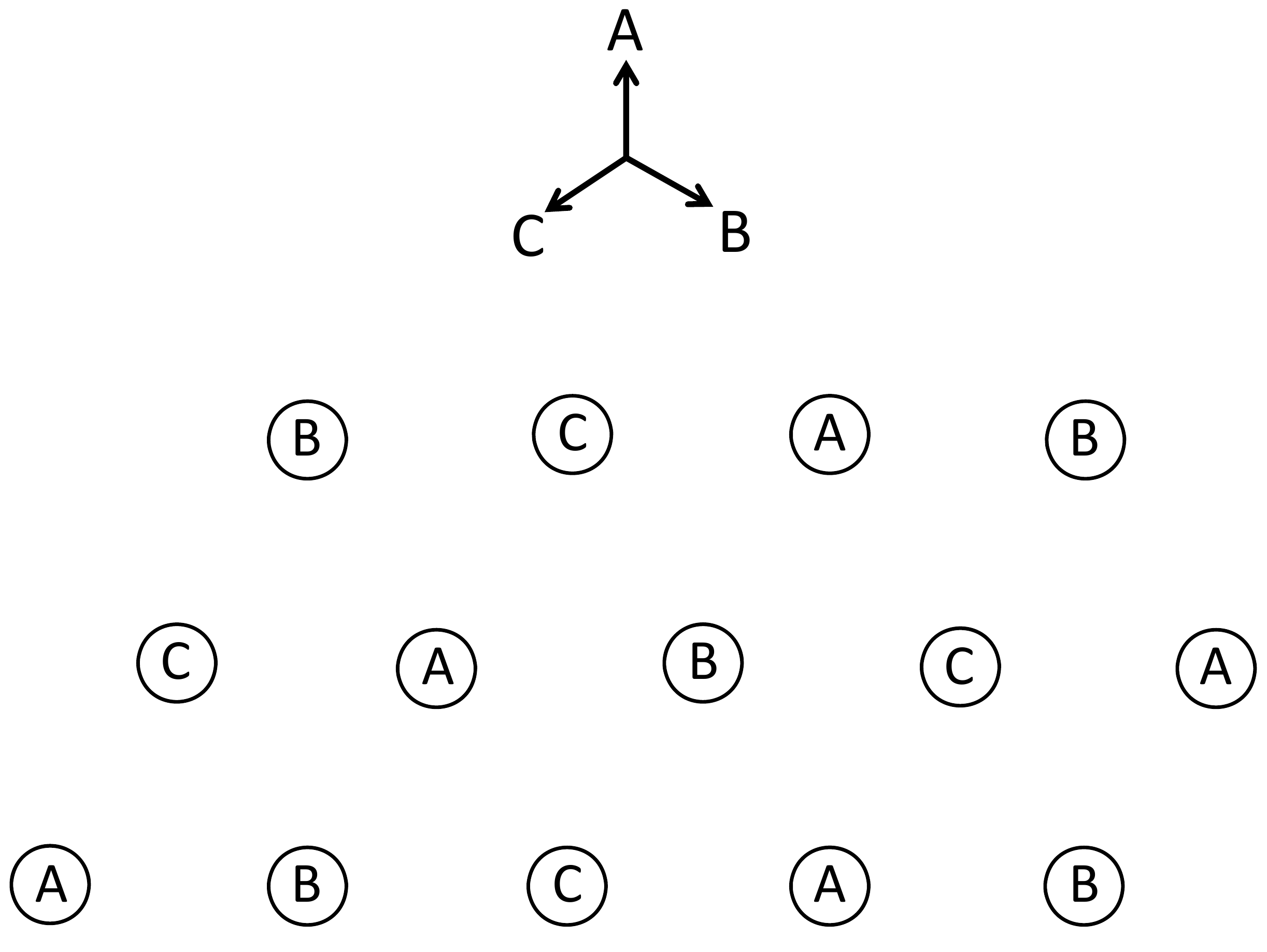}}\\
\end{tabular}
\caption{\label{fig:16}
 (Color online) Modulus of the spin susceptibility in real space at
  $T/J=0.375$ on the logarithmic scale.  Lattice points are enumerated
  according to their distance from the origin.  Spins on red sites (0, 2, 5, 6, 10)
  are correlated ferromagnetically while spins on blue and black sites are
  correlated anti-ferromagnetically with the spin at the origin.  At
  slightly higher temperature $T/J=0.5$ black points (3, 7)
  start to correlate ferromagnetically with the spin $S_0$ at the origin,
  contrary to the classical ground state pattern depicted to the right.
}
\end{figure}

It is clearly seen in Fig.~\ref{fig:15} that around $T/J=1$ system's
response is enhanced along the whole Brillouin zone boundary
indicative of the frustrated behavior. Only at temperatures below
$T/J=0.5$ it becomes evident that the system wants to develop
correlations commensurate with the K point.  We confirm previous
observation~\cite{singh} that even at $T/J = 0.375$ the spin
correlation length, which can be estimated from the half-width of the
peak around K point, is still of the order of lattice constant $a$.
This can be checked even more explicitly by looking on the static spin
correlations in real space.  Figure~\ref{fig:16} shows that while the
{\em sign} structure of short-range spatial correlations is consistent with the
three-sublattice $120^\circ$ state, the magnitude of the correlations
becomes exponentially small on the scale of a few lattice periods.  Closer look
also reveals dramatic suppression of correlations between site $0$
(say, sublattice A) and sites $3$ and $7$ both of which would belong
to sublattice C in the perfectly ordered $120^\circ$ classical state. Moreover,
at slightly higher temperature $T/J=0.5$ the sign of correlations on
sites $3$ and $7$ changes sign and turns ferromagnetic, similar to
A-sublattice spins, though with much smaller amplitude. This
temperature induced reversal of correlations is a remarkable effect
specific for a frustrated system.
%
%

It has been noted some time ago that short wavelength spin excitations
contribute significantly to the finite temperature properties of
triangular lattice antiferromagnet at not too low
temperature~\cite{zheng2006}. This has to do with substantial phase
space volume these excitations occupy as well as with their relatively
weak dispersion~\cite{starykh2006}. It is conceivable that
particularly weak correlations between sublattices A and C noted above
have to do with these excitations as well.  All of these features can
be extracted from the retarded spin susceptibility $\chi(q,
\omega)$ calculation of which requires analytic continuation of our
Matsubara-frequency susceptibility $\chi(q, \omega_m)$ to
the real frequency. We plan to address this important issue in the
near future.

\section{Conclusions}
\label{sec:7}
This paper describes novel approach to frustrated spin systems.
Obtained numerical results for the spin-1/2 triangular lattice Heisenberg
model, Section~\ref{sec:5}, show the power and competitiveness of our approach
in comparison with other well established numeric techniques.

Future work has to address the issue of performing simulations at
lower temperature in the regime characterized by the large correlation
length. Technically, this translates into being close to zero in the
denominator of Eq.~\eqref{dysonGW} for some momentum values.  Progress
in this direction should allow us to better describe the
cross-over/transition from the cooperative paramagnet to the
long-range ordered (albeit frustrated) state.

Yet perhaps the most promising line of attack has to do with applying
our technique to the geometrically frustrated models that do not
support magnetically ordered state at all.  In two dimensions this
singles out quantum kagom\'e lattice antiferromagnet which have
recently being shown to realize a long-sought $Z_2$ spin
liquid state~\cite{white2011}. This task will require extension of
our approach to systems with several (three in this case) spins in a
unit cell. The added matrix complexity does not represent any
fundamental difficulty.

Moving one dimension higher brings one to the most frustrated
antiferromagnet in the world -- spin-1/2 pyrochlore
antiferromagnet~\cite{canals98,moessner98,gardner2010}. Pyrochlore
Ising-like model realizes beautiful quantum spin-ice
physics~\cite{q-ice} while the fate of spin-1/2 Heisenberg model is an
open question. It is widely believed that `cooperative paramagnet'
region is most extended in this three-dimensional system.  Unlike many
lower-dimensional frustrated system, spin-1/2 pyrochlore is
essentially not accessible by quantum Monte-Carlo technique due to
large unit cell (4 spins).  We believe that our Diagrammatic MC
approach is therefore uniquely suited for studying the
finite-temperature dynamics of this outstanding frustrated magnet.

Recently we have generalized the Popov-Fedotov trick to a universal
technique of fermionization which leads to a well-defined standard
diagrammatic technique for arbitrary lattice spin, boson, and fermion
system with constraints on the on-site Fock
states~\cite{Fermionization}. This development creates a broader
context for the present work: successful implementation of the BDMC
method for models of quantum magnetism may lead to the universal
numerical tool for arbitrary strongly correlated lattice models within
the fermionization framework when diagrammatic expansion do not
involve large parameters.

We thank M. Rigol for communicating us data obtained within the NLC
method.  This work was supported by the National Science Foundation
under grants PHY-1005543 (S.K., N.P., B.S., and C.N.V.)  and DMR-1206774
(O.A.S.), and by a grant from the Army Research Office with funding
from the DARPA.

\bibliography{references}

\end{document}